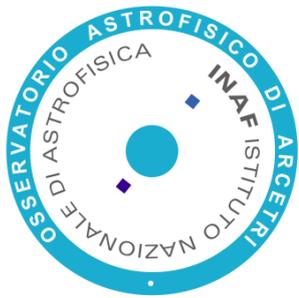

# Optical calibration of large format adaptive mirrors

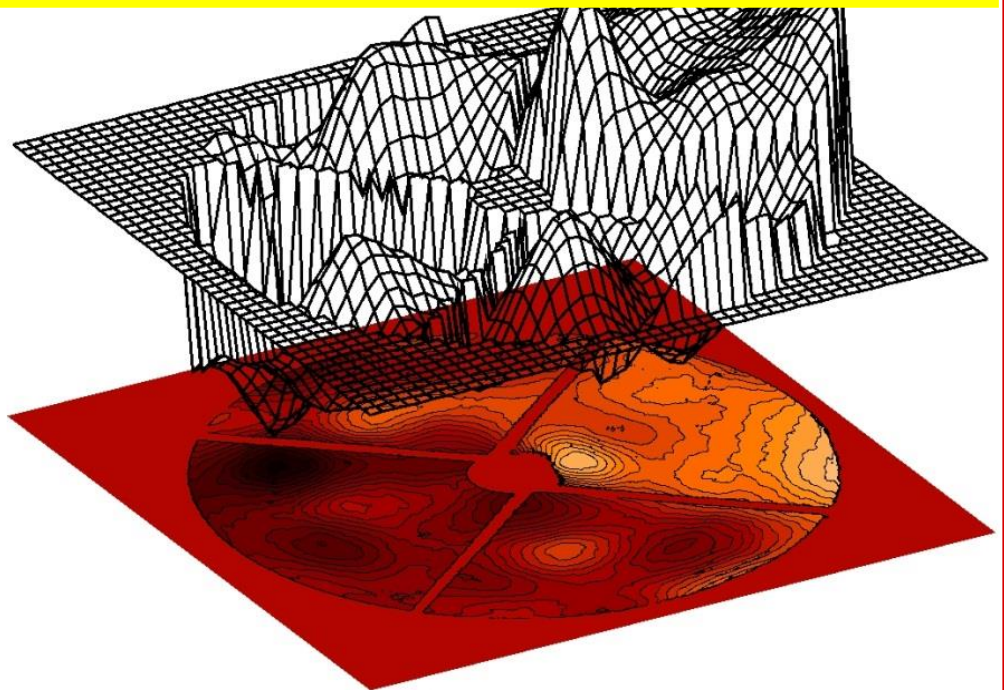

**Runa Briguglio**
**Marco Xompero**
**Armando Riccardi**

INAF – Istituto Nazionale di Astrofisica
Osservatorio Astrofisico di Arcetri

Front cover by Marco Xompero

# Table of Contents

# Abbreviations, acronyms and symbols

| Symbol | Description |
|---|---|
| AOF | Adaptive Optics Facility |
| ASL | Above sea level |
| CGH | Computer Generated Hologram |
| DC | Duty-Cycle |
| DSM | Deformable Secondary Mirror |
| DOF | Degree of Freedom |
| ESO | European Southern Observatory |
| INAF | Istituto Nazionale di Astrofisica |
| PTV | Peak to Valley |
| RMS | Root Mean Square |
| VLT | Very Large Telescope |
| WFE | Wavefront error |
| WRT | with respect to |

# References

[1] Del Vecchio, Ciro; Briguglio, Runa; Riccardi, Armando; Xompero, Marco: *Analysis of the static deformation matching between numerical and experimental data on the voice-coil actuated deformable mirrors*; **Proceedings of the SPIE, Volume 9148**, id. 914840 7 pp. (2014)

[2] Briguglio, Runa; Biasi, Roberto; Xompero, Marco; Riccardi, Armando, & al.: *The deformable secondary mirror of VLT: final electro-mechanical and optical acceptance test results*; **Proceedings of the SPIE, Volume 9148, id. 914845 8 pp. (2014).**

[3] Pariani, G., Briguglio, R., Xompero, M., et al.: *Approaches to the interferometric test of large flat mirrors: the case of the adaptive M4 for E-ELT*; **Ground-based and Airborne Telescopes V, 9145. 2014**

[4] Briguglio, R., Xompero, M., Riccardi, A., et al.: *Optical calibration and test of the VLT Deformable Secondary Mirror;* **Proceedings of the Third AO4ELT Conference, 2013**

[5] Briguglio, R., Xompero, M., & Riccardi, A.: *LBT adaptive secondary mirrors: chopping procedures and optical calibration on the test bench*; **Adaptive Optics Systems III, 8447. 2012**

[6] Briguglio, R., Xompero, M., Riccardi, A., et al.: *Optical calibration of capacitive sensors for AO: strategy and preliminary results*; **Adaptive Optics Systems III, 8447. 2012**

[7] Riccardi, A., Xompero, M., Briguglio, R., et al.: *The adaptive secondary mirror for the Large Binocular Telescope: optical acceptance test and preliminary on-sky commissioning results*; **Adaptive Optics Systems II, 7736. 2010**

[8] Del Vecchio, Ciro; Agapito, Guido; Arcidiacono, Carmelo; Carbonaro, Luca; Marignetti, Fabrizio; De Santis, Enzo; Biliotti, Valdemaro; Riccardi, Armando *The actuator design and the experimental tests of a new technology large deformable mirror for visible wavelengths adaptive optics,* **Adaptive Optics Systems III. Proceedings of the SPIE, Volume 8447, article id. 844708, 10 pp. (2012)**

[9] Biasi, Roberto; Gallieni, Daniele; Salinari, Piero; Riccardi, Armando; Mantegazza, Paolo, *Contactless thin adaptive mirror technology: past, present, and future,* **Proceedings of the SPIE, Volume 7736, id. 77362B (2010).**

[10] Xompero, Marco; Riccardi, Armando; Zanotti, Daniela, *Adaptive secondary mirror for LBT and its capacitive sensors: how can we calibrate them?*
**Adaptive Optics Systems. Edited by Hubin, Norbert; Max, Claire E.; Wizinowich, Peter L. Proceedings of the SPIE, Volume 7015, article id. 70153Q, 9 pp. (2008).**

[11] Riccardi, A*., Optical figuring specifications for thin shells to be used in adaptive telescope mirrors,* **Advances in Adaptive Optics II. Edited by Ellerbroek, Brent L.; Bonaccini Calia, Domenico. Proceedings of the SPIE, Volume 6272, id. 627250 (2006).**

[12] Riccardi, A., Xompero, M., Busoni, L.: *Fitting error analysis for the VLT deformable secondary*, **Advances in Adaptive Optics II. Edited by Ellerbroek, Brent L.; Bonaccini Calia, Domenico. Proceedings of the SPIE, Volume 6272, id. 62724O (2006)**

[13] Shin Oya, Aurelien Bouvier, Olivier Guyon, Makoto Watanabe, Yutaka Hayano, Hideki Takami,&al.: *Performance of the deformable mirror for Subaru LGSAO,* **Volume 6272: Advances in Adaptive Optics II**

[14] Anand Sivaramakrishnan, Ben R. Oppenheimer , *Deformable mirror calibration for adaptive optics systems*, **Proceedings Volume 3353: Adaptive Optical System Technologies, September 1998**

[15] Daniel Malacara, Optical Shop Testing, **ISBN: 978-0-471-48404-2**

[16] Jim Burge, Katie Schwertz, *Field Guide to Optomechanical Design and Analysis*, **SPIE Press Book, August 2012**

[17] James H. Burge, Chunyu Zhao, Ping Zhou, *Imaging issues for interferometry with CGH null correctors,* **SPIE Proceedings Volume 7739: Modern Technologies in Space- and Ground-based Telescopes and Instrumentation July 2010**

[18] James H. Burge, *Measurement of large convex aspheres*, **SPIE Proceedings Volume 2871: Optical Telescopes of Today and Tomorrow, March 1997**

[19] E. Molinari, D. Tresoldi, G. Toso, P. Spanò, R. Mazzoleni, M. Riva, A. Riccardi**,** R. Biasi, M. Andrighettoni, G. Angerer, D. Gallieni, M. Tintori, G. Marques: *The optical tests for the E-ELT adaptive mirror demonstration prototype,* **Proc. SPIE, vol. 7736, pp. 773632, 2010**

[20] D. Gallieni, M. Tintori, et al.: 2010, *Voice-coil technology for the E-ELT M4 Adaptive Unit,* **1stAO4ELT conf., EDP Sciences, 2010**

[21] Mark A. Ealey, John Wellman, "*Fundamentals Of Deformable Mirror Design And Analysis*"; **Proceedings Volume 1167, Precision Engineering and Optomechanics; (1989)**
[22] R. H. Freeman, J. E. Pearson, "*Deformable Mirrors For All Seasons And Reasons*"; **Proceedings Volume 0293, Wavefront Distortions in Power Optics; (1981)**
[23] P.-Y. Madec, "*Overview of deformable mirror technologies for adaptive optics and astronomy*", **Proceedings Volume 8447, Adaptive Optics Systems III; 844705 (2012)**
[24] James Kilpatrick, Adela Apostol, Anatoliy Khizhnya, Vladimir Markov, Leonid Beresnev, "*Real-time characterization of the spatio-temporal dynamics of deformable mirrors*", **Proceedings Volume 9979, Laser Communication and Propagation through the Atmosphere and Oceans V**
[25] Roberto Biasi, Daniele Gallieni, Piero Salinari, Armando Riccardi, Paolo Mantegazza, "*Contactless thin adaptive mirror technology: past, present, and future*", **Proceedings Volume 7736, Adaptive Optics Systems II; 77362B (2010)**
[26] Michael North Morris, Markar Naradikian, James Millerd, "*Noise reduction in dynamic interferometry measurements*", **Proceedings Volume 7790, Interferometry XV: Techniques and Analysis; 77900O (2010)**
[27] Neal Brock, John Hayes, Brad Kimbrough, James Millerd, Michael North-Morris, Matt Novak, James C. Wyant, "*Dynamic interferometry*", **Proceedings Volume 5875, Novel Optical Systems Design and Optimization VIII; 58750F (2005)**
[28] James Wyant, "*Precision interferometry in less than ideal environments*", **Proceedings Volume 7790, Interferometry XV: Techniques and Analysis; 77900J (2010)**
[29] B. Kimbrough, et al., "*The spatial frequency response and resolution limitations of pixelated mask spatial carrier based phase shifting interferometry*", **Proc. SPIE 85706(1), (2010)**
[30] M. Born and E. Wolf, "*Principles of Optics*", **(Cambridge University Press, 1999).**

# 1   Introduction

A deformable mirror does not have its own optical shape. You need to command it to your desired figure.
Within this scope, we will address the elements of the optical measurement, characterization and calibration of a deformable mirror. "Adaptive" or "deformable" are used here as synonyms, since deformable mirrors are used as wavefront correctors in adaptive optics systems. We consider as "large format" those mirrors whose size implies a complex, dedicated optical test facility to be measured: for instance, a telescope secondary or relay mirror. There is no conceptual difference between a 1 inch and a 1 m deformable mirror; both can be controlled and measured in the very same way. Conversely, the testing environment is a relevant difference, and the specific testing protocols to guarantee the successful completion of the calibration activities. Nanometer-level optical measurements are routine in a quite laboratory, while the noise background in an industrial facility may be significantly larger than the requested calibration accuracy. This is why we will describe here specific strategies to command the mirror and to assess the measurement noise, to identify the most suitable sampling parameters to achieve the wanted signal to noise ratio.

The context of this work is astronomical adaptive optics and more specifically the preliminary fine tuning of the wavefront corrector before closing the loop. A substantial part of such tuning consists in the calibration of instrument-specific auxiliary loops: Zernike commands for non-common path aberrations, fitting of KL modes to run the loop, e.g. We will not discuss them in detail since, as mentioned, they may be specific for a particular instrument.
Conversely, we try in the next section to clarify the "behind the scenes": actors, requirements, interfaces of a calibration procedure. Such aspects are in facts common to adaptive mirrors of all flavors; moreover a carefully design of the test setup (including requirements, trade-off, interfaces) is the first mandatory element to success.

This work is the output of the INAF grant TECNO-PRIN 2010, aiming at assessing "an automated calibration procedure" for adaptive mirrors. For this reason we discuss in deep the background and the procedural aspects, rather than the scientific or technological motivation; the results of each step are commented, in order to outline a sort of merit function and estimate the data quality. "Automation" was one of the main driver, thus justifying also the tabular format we adopted to describe the calibration steps.

In Sec. 2 we present the main concepts behind the measurement of a deformable mirror; in Sec. 3 we focus on the requirements of the various actors, to allow a large degree of automation in the optical samplings; in Sec. 4 we present a detailed list of measurement and analysis procedures. This part is organized with a common structure for each activity:
- a generic description to clarify the context;
- the data sampling procedure;
- an example of sampling parameters (referred to the peculiar case of the VLT-DSM, as explained in that section);
- the analysis procedure;
- the results, shown to identify a quality indicator for the measurement;
- an advanced analysis procedure, when some additional processing is requested.

We adopted such schematic, tabular approach for a state-machine view of the process. When we consider the next generation of adaptive mirrors, equipped with thousands of actuators, fully automated procedures will be a must to minimize the activity schedule, risk and cost and maximize the performances.

# 2 Conceptual elements of the optical calibration procedure

## 2.1 Optical calibration goal and overview

The calibration of a deformable mirror is requested to enhance the performances of an AO system and to provide a mirror flattening command. Such goals are obtained through the measurement of the mirror surface deformation produced by an external command or perturbation, thus improving the nominal characterizations provided by the design and FE models. The detailed calibration objectives are specified by the AO system top-level requirements and the calibration scheme will be naturally tailored on them. Within this scope, we report a wide band approach on the characterization of deformable mirrors, irrespective of their working principle and working conditions, to preserve a general validity; such preliminary description is requested to guide the preparation of the software package needed for the calibration. In the next chapters, we will then focus our attention more specifically on mirrors equipped with thin glass shell, contactless actuators and co-located internal metrology. The description of the mirror we tested will be given in Sec.4.1.

The mirror calibration is based on the measurement of its properties with an external feedback, which is usually provided by mean of optical metrology with an interferometer. The measured properties are the mirror surface deformations produced by given commands or perturbations. Two types of deformations and related measurements are commonly encountered: differential and absolute. For instance, differential sampling is requested to measure the mirror response to differential commands, i.e. irrespective of the initial mirror position; an absolute measurement is on the contrary needed to sample the actual mirror shape, e.g. for the flattening command accuracy verification. Both kinds of measurements are implemented with different techniques to improve the SNR as they are affected by a pretty different characteristic noise. The optical calibration takes place after the electromechanical characterization of the mirror, which is devoted to verify the full functionalities of the system and to the delivery of its relevant electromechanical properties. The mirror is then installed on the optical bench, composed by the interferometer, an optical system to image the mirror on the interferometer camera and a moving stage to provide the optical alignment.

## 2.2 The optical system

In order to image the mirror surface an optical system is requested; it is designed to magnify the optical beam to the mirror size and to optically couple it with the interferometer, obtaining an interference image.

The design of the optical system may be very demanding and its procurement very expensive, given the optical prescription of the deformable mirror and the accuracy requirements.

## 2.3 Conceptual steps of the calibration process

A deformable mirror, equipped with $n$ independent actuators, may be commanded by mean of a control matrix, which is a set of $n$ orthogonal vectors of $n$ actuator positions. Any mirror command may be then expressed as a linear combination of the control vectors. The mirror calibration is based on the optical measurement of the control matrix vectors; through this step, the internal metrology reading is referenced with the physical (optical) displacement of the glass, as measured by the interferometer. Any desired optical shape may be then computed by projection on the measured control matrix. Such point is broken down in a conceptual procedure, which may be summarized in the following points

- Optical alignment control
- Calibration of the internal metrology
- Mirror flattening and verification
- Case-dependent performances verification

Operatively, depending on the AO top-level requirements and on the properties of the deformable mirror system, these steps are implemented with different strategies.

### 2.3.1 Optical alignment control

The optical alignment is obtained by moving the motorized stages holding the deformable mirror and the optical elements on the bench. A periodic correction is necessary to compensate for thermo-mechanical drifts of the structures. The correction is performed by mean of a closed-loop between the optical shape as seen by the interferometer and the moving stages. The loop is initially calibrated by measuring the alignment Zernike aberrations produced by a known stage movement; then it is closed at the desired frequency or even upon user's command.

### 2.3.2 Calibration of the internal metrology

The internal metrology is calibrated by comparing a known command applied to the mirror with the optical result measured by the interferometer. Depending on the system complexity, such process may be iterated; also, the measurement strategy may be changed in order to span the metrology working range and enhance the accuracy; a sequence of additional steps may be requested to implement complex calibration strategies.

The direct measurement of the control matrix is the most straightforward technique for the calibration, at least for the initial stage; such matrix may be assembled by considering individual actuator movements, or the stiffness eigenmodes of the mirror, or any orthogonal combination of *n* actuator commands. The measurement is based on a differential sampling because a differential command is applied (i.e. it is irrelevant the initial position); each command will be sampled *p* times to improve the measurement SNR; the duration of this step is therefore the most time consuming step within the test schedule.

### 2.3.3 Mirror flattening and verification

The mirror shaping (or flattening, as an optically null shape is typically requested) is obtained by combining the control matrix vectors to produce a null signal on the interferometer; the obtained shape is then measured to qualify the result. In this case, an absolute sampling is requested. To improve the measurement accuracy the convection disturbance, systematic errors and quasi static noise must be controlled.

The mirror flattening procedure may be iterated as long as the control matrix is refined.

### 2.3.4 Case-dependent performances verification

Depending on the AO requirements, a sequence of tests may be scheduled to verify some specific system performances. Many dedicated tests may share the same sampling mechanism: with such strategy it is possible to re-use existing items like software packages, errors control procedures, noise analysis. A list of case dependent verifications may include:

- Temporal stability of the optical shape
- Temperature dependence of the optical shape
- Verification of the FEA model
- Field stabilization and chopping command verification
- Calibration of the AO control basis (e.g. Karhunen-Loève modes)

## 2.4 Subsytems and interfaces

So far, we identified the subsystems involved in the calibration process:

- The deformable mirror
- The interferometer
- The alignment stages
- The software package and the workstation

Such subsystems are linked together by mean of interfaces, whose task is to allow the flow of measurement data, status information and commands among the subsystems. The identification of the interface properties is requested to address the hardware requirements and drive accordingly the subsystems design.

| From | To | Data | Interface |
|---|---|---|---|
| **Mirror** | Interferometer | Synchronization trigger | Electrical/optical |
| | Alignment stage | - | Mechanical |
| | Software | Diagnostic data<br>Status indicator<br>Configuration | Digital communication |
| **Interferometer** | Mirror | - | No interface |
| | Alignment stage | - | No interface |
| | Software | Images<br>Status indicator<br>Configuration | Digital communication<br>Data archive access |
| **Alignment stages** | Mirror | - | Mechanical |

|  | Interferometer | - | No interface (case dependent: interferometer optics alignment tools) |
|---|---|---|---|
|  | Software | Position<br>Status indicator | Digital communication |
| **Software & Workstation** | Mirror | Commands<br>Command buffer<br>Configuration | Digital communication |
|  | Interferometer | Commands<br>Configuration | Digital communication |
|  | Alignment stage | Commands<br>Configuration | Digital communication |

Table 1

## 2.5 Data organization

Each step of the calibration will produce a certain amount of measurement data, which will be used to perform a following procedural step or will be analyzed to produce a test result. The sampling parameters and the system configuration will be also saved together with the data acquired; a general scheme of the data to be saved and their format is given in Table 2. An efficient data organization is required to allow their automated management through the analysis pipeline: a conceptual scheme useful for data naming and organizing is given in Table 3 and Figure 1.

Files and folders name are coded in a configuration file: this allows preserving the functionalities of the acquisition and data analysis software when changing workstation, or when processing remotely the data. This feature allows also using the same software package for the testing of different deformable mirrors, or in the case of a segmented system, where the full mirror is composed by sub-mirrors to be tested individually. This is the case, for instance, of the E-ELT M4 mirror or of the GMT secondary.

A preferable data format is the FITS, as it can be easily read in any platform; also, the keywords that are saved in the FITS header are attractive to store additional informations regarding the data (e.g. date/time, configuration version, software version). As the FITS format allows also storing multidimensional data, we adopted the packing convention in Table 2.

| **Data type** | **Example** | **Packing** |
|---|---|---|
| scalar (float/double) | temperatures; # modes to be corrected | When possible, scalars are grouped together in vectors according to a template; otherwise, each scalar value is saved in a single FITS file |
| lists (long) | disabled actuators; marker actuators; valid points within an image. | saved as column vectors in a FITS file. |
| 1 D vectors (float/double) | Hexapod coordinates; command amplitude; | saved as column vectors in a FITS file. |
| 2D vectors (matrices, float/double) | FF matrix; command matrix; actuator coordinates | saved as 2 D vectors in a FITS file. |
| 2D vectors (images, float/double) | interferometer images, e.g. the flattening result. | Saved as 3D vectors in a FITS file; the first plane is the image, the second one is the associated mask. |

| 2D vectors (images group, float/double) | images group, e.g. optical interaction matrix, selection of the points within the intersection mask | Saved as a 2D vector, where each column contains the valid pixels of each single image. |
|---|---|---|
| 3D vectors (images group, float/double) | interferometer group of raw images, e.g. the optical interaction matrix | saved as 3D vectors in a FITS file; the first plane is the image, the second one is the associated mask. |

Table 2

### 2.5.1 Meta data saving

Together with the samples that are acquired during each measurement, there is a large amount of data that are needed to describe the system or the sampling configuration. Such additional data are used for three purposes:

- To qualify the measurement procedure, i.e. to keep records of the sampling parameters, as they might be used at a later stage of the test to post-process the associated data. This is the case, for instance, of the applied command amplitude when measuring the mirror influence functions: the image corresponding to each mirror mode will be normalized to the actual commanded amplitude.

- To qualify the external conditions of the sampling, for instance recording the cooling temperature, the stages alignment positions.

- To record the mirror status at the time the measurements, for instance actuator positions and forces, control matrix, working actuators.

This is not only needed for a general qualification of the datasets but also to restore the system configuration to be used at a later time.
The system configuration is changed upon user's request only: it follows that modifications occur at a very much slower time scale with respect to the data sampling; this means that the same configuration will be used during many days of testing activity. In order to optimize both disk usage and the tracking number data organization, we adopted for the metadata the following mechanism: when a dataset is created with its own tracking number X, a new link X is generated in the configuration folder; the link destination is the last saved configuration.

### 2.5.2 The tracking number

The tracking number is an identification string that labels each datum. We adopted a time based tracking number mechanism, so that whenever a new datum is produced a new tracking number is generated, in the form year-month-day_time, e.g. 20130101_123400. The tracking number string is used to name the data destination folder: inside that folder, the FITS files are named according to the variable they contain. As the filenames are identical inside each folder, it is possible to analyze them transversally: this is the case, for instance, of analyzing the temporal behavior of a given variable.

Tracking numbers may be grouped together as array: this feature is requested for instance when dealing with datasets that are contained in many folders, as they have been sampled at different times, and must be put together for processing. A typical case is that of the mirror influence function (see 4.2), where a first folder contains the first n images, the second folder contains the next n and so on.

We adopted the convention of tracking number fatherhood, meaning that a new tracking number is created only when a brand new dataset is generated; when data are analyzed and a result is produced, the tracking number of the results is the one of the dataset they have been produced with. Such concept allows keeping track of the data processing path, moving from one folder to the next with the same label. When tracking numbers are used in array form, the first in the array is the one passed to the result.

The limitation of such mechanism is that it doesn't allow re-processing the data (data will be over-written when processed the second time). This is a wanted feature, as the only case when it would be desired to re-process the data is after modifying the software.

### 2.5.3 Versions repository

The versions repository tool is adopted to keep track of the software modifications and of the system configuration parameters. A commercial package was used within this scope.

| Acquired dataset | Assoc. metadata | Proc. dataset | Folder name | Tracking number | Content |
|---|---|---|---|---|---|
| System configuration | System configuration | | | **20120101_000100** | FF_matrix.fits, Command_matrix.fits, good_actuators.fits |
| Interferometer images for the IF sampling | | | Data/Phasemaps/ | **20120101_010000** 20120101_010100 20120101_010200 | dataset: mode#0 to mode#250 dataset: mode#251 to mode#500 dataset: mode#501 to mode#750 |
| | System configuration | | Conf/ | 20120101_010000 20120101_010100 20120101_010200 | FF_matrix.fits, Command_matrix.fits, good_actuators.fits |
| | Sampling parameters | | Data/Mirsetup/ | **20120101_010000** | command amplitude, command matrix, #of sampled modes, # push-pull |
| | | Noise evaluation | Result/ Phasemaps/ | **20120101_010000** | Structure_function.fits; Tip_tilt_vs_num_frames.fits HighOrder_vs_num_frames.fits |
| | | Reconstructed mirror modes images | Data/If_functions | **20120101_010000** 20120101_010100 20120101_010200 | dataset: mode#0 to mode#250 dataset: mode#251 to mode#500 dataset: mode#501 to mode#750 |
| | | Mirror Int. Matrix | Data/IntMat/ | **20120101_010000** 20120101_010100 20120101_010200 | IM.fits IM.fits IM.fits |
| Reference image and position for flattening | | flattening result | Data/Flat/ | 20120202_010000 | start_image.fits, flat_image.fits, flat_command.fits, averaged_time.fits |
| | System configuration | | Conf/ | **20120202_010000** | FF_matrix.fits, Command_matrix.fits, good_actuators.fits |
| | | | Result/Flat | **20120202_010000** | Flat-4Zern.fits Flat-10Zern.fits Analytical_flat.fits ForcePattern.fits |
| Interferometer images for capsens calibration | | | Data/Phasemaps/ | **20120303_010000** 20120303_010200 20120303_010300 | Images collected during run #1 Images collected during run #2 Images collected during run #3 |
| | System configuration | | Conf/ | 20120303_010000 20120303_010100 20120303_010200 | FF_matrix.fits, Command_matrix.fits, good_actuators.fits |
| | Sampling parameters | | Data/Mirsetup/ | **20120303_010000** 20120303_010100 20120303_010200 | command amplitude, command history, # push-pull |
| | | Reconstructed images | Data/CapsensCal | **20120101_010000** 20120101_010100 20120101_010200 | dataset: run #1 dataset: run #2 dataset: run #3 |
| | | Capacitive sensor calibration results | Data/ActCalib/ | 20120303_010000 | |
| Hexapod Int mat | | | Data/HpIntMat | 20120101_020000 | image_cube.fits, HP_pos.fits, HP_IM.fits |

Table 3 Data organization table

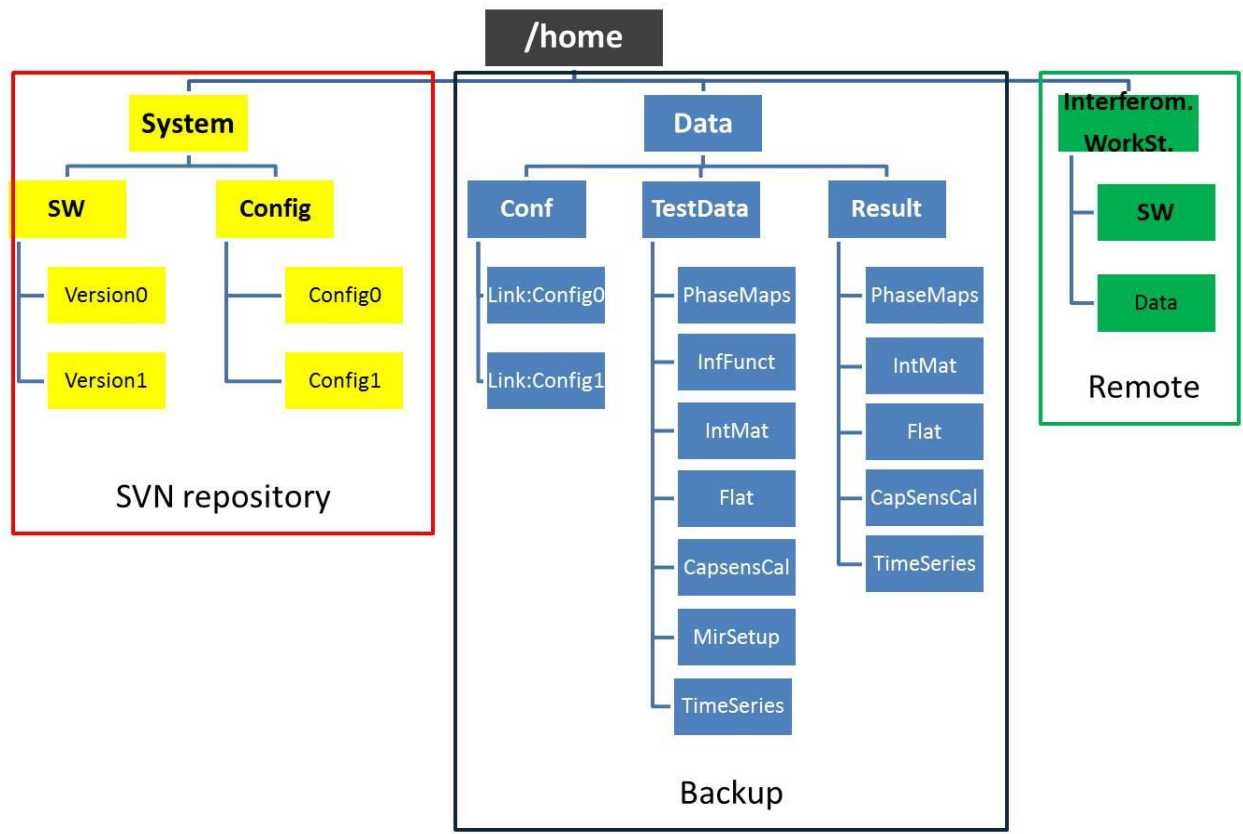

Figure 1 Data organization chart

## 2.6 Optical calibration timeline

In Sec. 2.2 we reported the generic steps of the calibration procedure; we will try now to break down those conceptual points into an activity schedule. The goal is to identify the process dependencies and prerequisites and the subsystems status during the full procedure. As the process steps produce a result and a modification of the system status, it is possible to identify process indicators or checkmarks: they can be used as criteria for the test progressing. In Table 4 a hypothetical activity schedule is given.

| # | Status in | Activity | Output | Status out | Checkmark |
|---|---|---|---|---|---|
| 1 | separated subsystems | subsystems installed on optical bench |  | integrated system |  |
| 2 | integrated system | optical alignment | alignment position | integrated system | interferometer image |
| 3 | integrated system | preliminary mirror shaping | mirror command M0 | mirror shape S0 | pupil visibility |
| 4 | mirror shape S0 | alignment closed loop calibration | alignment parameters | mirror shape S0 |  |
| 5 | mirror shape S0 | control matrix measurement | control matrix data | mirror shape S0 |  |
| 6 | mirror shape S0 | preliminary flattening | mirror command M1 | mirror shape S1 | WFE |
| 7 | mirror shape S1 | control matrix measurement (iter #2) | control matrix data | mirror shape S1 |  |
| 8 | mirror shape S1 | preliminary flattening (iter #2) | mirror command M2 | mirror shape S2 | WFE |
| 9 | mirror shape S1 | full flattening | mirror command M3 | mirror shape S3 | WFE |
| 10 | mirror shape S3 | metrology calibration (iter#1) | calibr. parameters | mirror shape S3 |  |
| 11 | mirror shape S3 | metrology calibration (iter#2) | calibr. parameters | mirror shape S3 |  |
| 12 | mirror shape S3 | control matrix measurement | control matrix data | mirror shape S3 |  |

| 13 | mirror shape S3 | final flattening | mirror command M4 | mirror shape S4 | |
| 14 | mirror shape S4 | time series test | stability specs | mirror shape S4 | |
| 15 | mirror shape S4 | field stabilization test | field stab specs | mirror shape S4 | |
| 16 | mirror shape S4 | chopping test | chopping specs | mirror shape S4 | |
| 17 | mirror shape S4 | temperature variation test | temperature specs | mirror shape S4 | |

Table 4

# 3 System functional requirements

In order to implement the interfacing mechanism described in Sec. 2.4, each of the subsystems shall satisfy a number of requirements. Within this section we will present them as guidelines for the procurement or the design of the subsystems.

## 3.1 Deformable mirror requirements

### 3.1.1 Full functionality

At the end of the electromechanical test, before installation on the optical bench, the deformable mirror shall be tested to demonstrated full functionality and long term stability. Such requirement is intended to allow the successful completion of the optical test without failures on the system.

In particular, a dry run of the optical test routines shall be executed to verify that the mirror is able to apply all the wanted command without any failure.

### 3.1.2 Digital communication port

The mirror electronics shall be equipped with a digital communication device; it is accessible by the supervising workstation and software on the same local network. The link is fast enough to allow bulk data transfer (for instance: 1k actuators, position and force sampled at 1kHz). The mirror configuration and commands are sent from the workstation through this link.

### 3.1.3 Status tracing

The low level system status shall be accessible upon request. Such requirement is intended to allow the progressing of the automated routines when a status check is passed. The system status, for instance, may be listed as:

- System powered on/off
- General failure
- Actuators enabled/disabled
- Command applied/discarded

### 3.1.4 Internal diagnostics

The relevant diagnostics information shall be retrievable from the mirror electronics, where they are stored in a buffer containing a diagnostic time history. The sampling frequency shall be configurable.

### 3.1.5 Internal clock and counter

The mirror electronics shall feature an internal clock running at a configurable frequency (e.g. 100 Hz to 1 kHz). Such clock is used as a time reference for the application of commands and for saving the diagnostic data. A progressing counter is used for the events time alignment.

### 3.1.6 Command sequence buffer

The mirror electronics shall be equipped with an internal buffer, where a sequence of commands may be stored upon request. The commands are applied at the internal clock frequency, i.e. a command is applied when the mirror counter is incremented by one step.

### 3.1.7 Trigger line

The mirror electronics shall be equipped with a trigger port. A signal (TTL pulse, optical pulse) is output from such port with the internal clock. The port shall be enabled/disabled to start/stop the triggering mechanism; a decimation factor and a delay factor between the port and the clock may be configured.

## 3.2 Interferometer requirements

In the following we will focus our attention on the requirements that are most specifically suitable for automatic operation, rather than general optical or mechanical specifications.

### 3.2.1 Dynamic interferometry

The instrument work principle shall be the dynamic interferometry, that allows collecting on a single shot the 4 images requested for the phase reconstruction. Such specification is critical: the typical exposure time on a dynamic interferometer is lower than 100us, so that convection and vibrations are freezed and the phase is correctly reconstructed.

### 3.2.2 Capture range

In dynamic interferometry a modulation carrier (a tilted mirror or a pixelated mask) is inserted in the internal optical beam. The carrier acts as a reference tilt to demodulate the phase signal measured on the CCD: the capture range depends on the CCD resolution and the tilt amplitude, which is typically tailored to match the Nyquist limit on the CCD. The capture range obtained in this way is by design larger than 10-20 um, which is a suitable requirement for the optical test procedures.

### 3.2.3 Frame rate

In order to consider the convection as a static noise in differential sampling, the interferometer frame rate shall be faster than the time when a convection structure crosses a resolution element in the image. Typically, a frame rate faster than 20 Hz is enough to reduce the convection noise down to a few nm residual error.

### 3.2.4 Remote communication

The interferometer software shall feature a interface for the remote communication with the supervising workstation and software: for instance, a Client-Server mechanism. The acquisition and configuration commands shall be executable remotely. Such request is intended to allow a fully automated interfacing between the interferometer and the mirror.

### 3.2.5 Scripting

The relevant SW commands that can be executed graphically from the GUI should be executable via command line on a dedicated console or by remote call.
A full list of the implemented commands must be provided, including a general description about building upper level scripts starting from the low level commands. Generic system commands for files management (e.g. copy, move, listing) should be callable.
The scripts provided should include at least (e.g.):
- A command to grab n raw frames and store them in the disk
- A command to post-process the a list or a folder of raw frames
- A command to enable/disable the synchronization trigger
- A command to set the image resolution
- A command to load a given configuration file
- A command to set the acquisition delay

### 3.2.6 Trigger line

The interferometer shall be provided with a synchronization mechanism to grab an image upon a certain condition. Such device may be for instance a trigger input line (with TTL standard).
The trigger mechanism must be such that a single image is captured when a single pulse is received (this, to specify that it is excluded the mechanism by which a sequence of images is captured at the internal frequency when the first trigger pulse is received).
The number of collected images must be equal to that specified in the command; if the trigger signal is stopped before the collection of the number of images requested, the acquisition is automatically stopped after a timeout with no error.
When the external trigger port is enabled, the system waits for the trigger signal during a start-of-trigger timeout. The timeout shall be in the range 5s to 10s. If no signal is received the system returns an error.
The slowest accepted frequency is given by the definition of the start of trigger timeout. When no trigger signal is received within the timeout, an error is returned.
The fastest triggering frequency is given by the frame grabber frame rate $f$. When a faster trigger signal is received, an image is acquired when a pulse is received after a time $t=1/f$ after the last captured image.
The synchronization mechanism must be not affected by the frequency stability, i.e., no stable or unique frequency must be expected (example: 100 images are requested; the first 50 are triggered with a 20Hz signal, then a 300ms delay is commanded by suspending the trigger signal, then the last 50 images are triggered with a 15 Hz signal)
The triggering mechanism can be configured with a delay $t$ between the detection of a trigger edge and the collection of an image. Such delay should be selectable in the range $0ms<t<1/2f$ (alternatively: $0ms< t<8ms$). The delay resolution must be comparable to the CCD exposure time (tens of microseconds). The delay must be selectable with a scripting command.

## 3.3 Optical bench requirements

### 3.3.1 Mechanical insulation

The optical bench shall be mechanically decoupled from the ground and/or the rest of the laboratory, to reduce the impact of vibrations in the optical test. The structure may be for instance mounted on damping elements.

The most annoying vibration component comes from high frequency (1kHz, e.g.), comparable to the interferometer integration time: the effect is that the instantaneous tilt changes during the image integration, producing an error in the phasemap reconstruction; the dampers should then be designed/procured to completely reject the high frequency oscillations.

### 3.3.2 Optical beam insulation

The volume in which the optical beam propagates shall be protected to avoid air blowing within the measurement arm. Such protection may be provided by a physical shielding with some lateral panels, for instance; they should be placed far enough from the optical beam to avoid laminar air flow through the beam itself.

The panels should also be thermally insulated to avoid propagating temperature variations to the test beam and cause then convection.

### 3.3.3 Thermal control

The mirror electronics is thermally controlled at a given temperature. To avoid that convection is excited by any vertical negative temperature gradients, the optical bench should be thermally controlled accordingly. The thermal design should follow the test geometry: if we assume that the mirror is suspended on top the optical bench, a positive vertical gradient may be created to stop convection. Such gradient is produced by setting the bench top to a temperature warmer than the mirror, and the bench bottom to a colder one. The vertical gradient should be maintained also on the lateral walls.

The bench temperature shall be monitored and actively controlled with a suitable remote command.

### 3.3.4 Additional case dependent requirements

Depending on the test goals, some additional requirements may be requested to the optical bench subsystem; they are intended to produce specific test conditions in order to calibrate the mirror behavior in a sort of simulated environment. They are listed in Table 5.

| Requirement | Task |
| --- | --- |
| Temperature control | To simulate night-time or seasonal temperature variation in the dome and calibrate the mirror accordingly. |
| Humidity control | To simulate the mirror behavior under different air humidity conditions. |
| Pressure control | To simulate the mirror behavior under low pressure (high altitude) conditions. |
| Elevation control | To simulate the mirror behavior when the telescope is moving along the elevation axis. |

Table 5

## 3.4 Optical test setup requirements

The optical setup shall be designed in order to allow differential or/and absolute calibration of the deformable mirror surface shape. For instance, the system is usually designed such as the test beam impinges normally the test surface, and the reflected beam is returned to the interferometer in a double-pass configuration.

In the case of deformable mirrors, the system shall allow the measurement of its surface shape in the nominal configuration (flattening) or when different shapes are applied (shaping). Actually, the system is usually optimized for the nominal configuration, and the overall error budget shall be limited to values well below the accuracy required for the calibration of the deformable mirror. This value shall be in the order of 0.03-0.05 waves RMS, i.e., 20-30 nm RMS.
When the mirror has to be shaped, the optical answer shall be as linear as possible with the mirror deformation, and retrace error, arising when the forward and backwards beam follow different optical paths, shall be taken into consideration.

Two are the main components in the overall error budget: errors coming from the optical components and errors coming from their incorrect alignment into the cavity.

### 3.4.1 Manufacturing errors

When optical elements are introduced into the interferometric cavity to shape the testing beam, their manufacturing errors may propagates into the cavity as a residual WFE. Considering for example a mirror with a surface defect on the surface with a deviation S, the reflected wavefront will be affected by an error of entity $d = 2S$. A transmission lens, in the same manner, will affect the wavefront with an error of $d = -(n-1)S$, where n is the refractive index of the substrate material. Furthermore, if the same component is used in double pass, its errors account twice on the cavity wavefront.

Since the deformable mirror can compensate for any WF deformation, absolute measurements are affected by such errors. The main strategies to limit their contributions consist in:
- o  maintain manufacturing errors under a threshold value, that is usually in the order of 1/10 of the calibration accuracy of the deformable mirror. This is achievable with the technology nowadays available, with a loss in an increasing of manufacturing costs.
- o  define a calibration strategy to subtract the residual error from the cavity. This can be usually performed when the size of the cavity is limited, and certified reference optics available for the calibration.

### 3.4.2 Alignment errors

The small misplacement of each optical element inside the interferometric cavity may introduce residual WFEs. The misplacement may be due to a wrong placement of the element as respect to its nominal position given by thermal dilatations or vibrations at low frequency of the optical elements. The sensitivity of the cavity to the misplacement of the different components shall be deeply studied, to identify the misplacement upper limit. Usually, some alignment errors are tolerated unless the induced wavefront error overcomes 0.01-0.02 waves RMS, i.e. 5-10 nm RMS.

# 4 The calibration procedure

In the following we will consider as a template the case of the VLT-UT4 deformable secondary mirror (DSM) on its test-bench names ASSIST. More information about these systems may be found in [4].

## 4.1 The DSM on ASSIST

The DSM is the deformable secondary mirror equipping the adaptive optics facility of the VLT UT4. The system is composed by a 1.2m diameter, 2mm thickness glass shell which is controlled by 1170 contactless, voice-coil actuators. Each actuator is equipped with a co-located capacitive positions sensor: an internal diagnostic loop is closed on the capacitive sensor position reading.
The mechanical support for the mirror is provided by an aluminum cold plate, which is also the connection point for the actuators. The cold plate holds a glass body acting as the capacitive sensor reference.
The DSM is installed on ASSIST, the ESO AOF optical test facility; ASSIST is equipped with a dynamic, Twyman-Green 4D Technology interferometer and a beam expander composed by a 1.6m large, concave AM1 mirror and a 12cm diameter convex AM2 one. The DSM is aligned within the optical bench by mean of a hexapod.

## 4.2 Preliminary flattening

Given the mechanical property of a glass thin shell, its optical shape depends on the supporting structure and on the actuator command applied. At the very beginning of the optical test, the optical shape of the shell is given by the actuators differential calibration errors and by the polishing and thinning residuals of the shell back surface and of the reference plate. The first image is then expected to be highly aberrated: an example is given in Figure 2 (left panel), where the local deformation in many areas is so large to be outside the interferometer capturing range.
Such initial condition is not suitable for starting any automatic calibration procedure. In particular, we considered the optical mask as a process indicator; under these initial conditions, the optical mask is so poor that the calibration process is not guaranteed to converge.
We tested a partially automated procedure, aimed at moving the mirror shape within the interferometer capturing range, i.e. to correct the very steep local deformations. The procedure is divided into two parts: in the first step (visibility improvement), the most critical areas are identified on the mirror and a user defined shape (e.g. a Gaussian deformation) is applied to correct them; in a second step (low orders correction), when most of the optical pupil is visible, the first mirror modes are applied sequentially with user-defined amplitude, trying to minimize the corresponding Zernike aberrations as measured by the interferometer; as the Zernike tip-tilt, piston and astigmatism are likely mixed together in the first 5 mechanical mirror modes, such step works better and faster when considering only modes from the 6$^{th}$ to the 11$^{th}$ (trefoil to pentafoil, e.g.).
Although such procedure is conceptually a trial-and-error process, it resulted an efficient way to speed up the initial stages of the calibration. A detailed description of the two steps is given in Table 6 and Table 7.

**Procedure: visibility improvement**

| Step | Action | Result |
|---|---|---|
| 1 | The interferometer is monitored in "Live Mode" on the workstation screen. | |
| 2 | The mirror engineering GUI displaying the actuator commands is monitored. | |
| 3 | The orientation angle/flip between the interferometer monitor and the GUI is identified. | GUIs relative orientation |
| 4 | The actuator coordinates x,y are restored from the suitable configuration file. | Actuator coordinates |
| 5 | A mirror area with a large local deformation is selected. The actuator $i$ located at the center of the spot is identified. | |
| 6 | A Gaussian command is computed, with equation: $C = A\ exp(\ -((x-x_i)^2+(y-y_i)^2)\ /\ 2s^2\ )$, where A is the amplitude of the desired deformation and s its spatial extension. | Gaussian command $C$ |
| 7 | The command C is applied. | Mirror shaped according $C$ |
| 8 | From the interferometer monitor, the user checks if the local slope has decreased. | |
| 9 | The parameters A and s are modified. | Modified $C$ command |
| 10 | Steps 6 to 8 are repeated until a sufficient correction is obtained. | |

| 11 | Steps 5 to 8 are repeated changing the considered area/central actuator. | |

Table 6

**Procedure: low orders correction**

| Step | Action | Result |
|---|---|---|
| 1 | The interferometer is monitored in "Live Mode" on the workstation screen. A "Zernike analysis" interface (which is a common alingment tool provided by interferometer software) is opened. | |
| 2 | The mirror modes matrix file is loaded and the first modes are considered. They are displayed to identify those matching trefoil to pentafoil (e.g.) | Mirror modes matrix |
| 3 | The measured amplitude of the Zernike trefoil (e.g.) is noted from the interferometer software. | Amplitude of Zernike modes from the interferometer. |
| 4 | The first considered mode (e.g. first trefoil) is applied with arbitrary amplitude. The corresponding measured values are checked. | Mirror shaped with the application of first mirror mode. |
| 5 | Steps 3 and 4 are repeated until the measured values are minimized. | |
| 6 | Steps 3 to 5 are repeated with the rest of the considered modes. | See Figure 2 |

Table 7

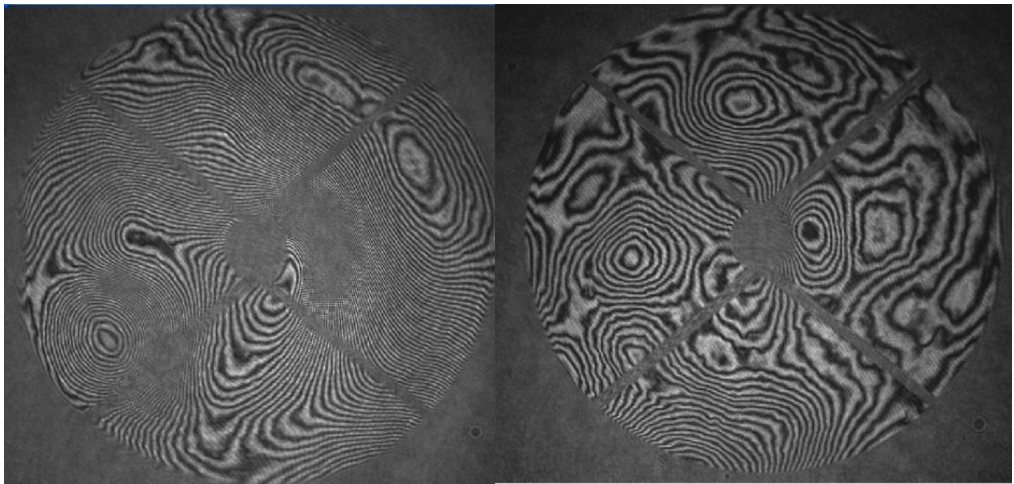

Figure 2 First image of the DSM thin shell as seen by the interferometer on ASSIST.

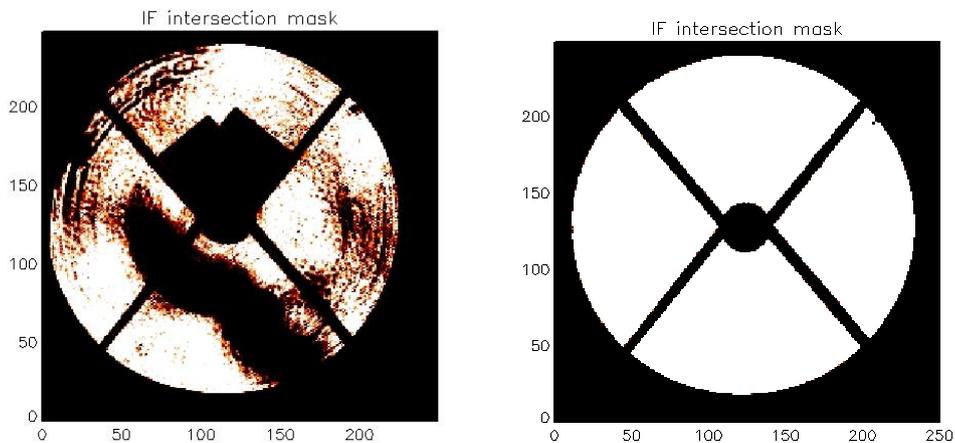

Figure 3 Left panel: IF intersection mask at the beginning of the flattening process, when the initial mirror deformation is so large that many areas are outside the interferometer capturing range. Right panel: IF intersection mask at the end of the flattening process; the pupil is visible and the full set of mirror modes can now be collected.

## 4.3 HP automated alignment procedure

The correction of the HP position is requested to preserve the optical alignment during the test. Such correction may be request frequently, depending on the mechanical stability of the optical system and on the steadiness of the environment conditions. An automated procedure is helpful to speed up the process and to allow long duration, unattended activities. The automated procedure is based on the knowledge of the coupling of the HP movements with the alignment aberrations as measured by the interferometer; such information may be retrieved from the optical design or measured directly on the optical bench. In the end, a projection matrix $Z_{HP}$ is obtained. At any time, the HP movement to null the alignment aberrations may be computed as the projection of the current aberrations on the $Z_{HP}$ matrix.

The procedure must be run when the pupil visibility is good enough to allow a decent Zernike measurement on the image; it is not however necessary to have a full visibility as we are interested in the measurement of very low order modes which may be estimated with a reasonable precision even on a partially visible pupil. The procedure may be then iterated as soon as the pupil visibility is improved.

The typical time scale of the correction must be also estimated, to implement a scheduled alignment during the optical test.

### 4.3.1 Data sampling

The sampling mechanism is based on the concept of interaction matrix: a known command is applied and the resulting effect is measured. The commands to be applied are given in the HP degrees of freedom space and they are: displacement in x, y and z, x rotation and y rotation. The resulting effects to be measured are the Zernike aberrations produced by the out-of-alignment position. The Zernike modes we are interested in are tip/tilt, power and coma; the higher order aberrations coming from the alignment may be neglected as their contribution is important (depending on the optical design) only far from the nominal alignment position.

Tip/tilt is coming both from alignment and vibration of the optical bench; this is the reason why it is necessary to collect time-averaged data in order to reduce the contribution of vibrations. The sampling procedure for the $Z_{HP}$ matrix is given in Table 8.

**Automatic sampling procedure: HP-Zernike coupling parameters**

| Step | Action | Result |
|---|---|---|
| 1 | The shell is set and a preliminary alignment position is found. | |
| 2 | A set of Zernike alignment aberrations is considered | Zernike alignment modes $Z=[1,2,3,4,7,8]$ |
| 3 | A vector $V$ of HP delta movements is considered: V=[x,y,z,Rx, Ry]. | |
| 4 | The amplitudes of the elements in $V$ is selected, according to current DSM fringes density, in order to not saturate the interferometer at any command in $V$. | Vector of HP commands: $V_{HP}=aV$ |
| 5 | A suitable image integration time is selected, according to the current alignment noise. | Image integration time $t$. |
| 6 | A movement element in $V_{HP}$ is applied. | HP moved in $V_{HPi+}$ |
| 7 | An image is collected with integration time t. | Image $s_{i+}$ of the alignment aberrations corresponding to $V_{HPi+}$ |
| 8 | The HP is moved to -$V_{HP}$ with respect to the starting position. | HP moved in $V_{HPi-}$ |
| 9 | An image is collected with integration time t. | Image $s_{i-}$ of the alignment aberrations corresponding to $V_{HPi-}$ |
| 10 | The differential image is computed and its $Z$ content is measured. | Z Zernike decomposition of $s_i = (s_{i+} - s_{i-})/2$ |
| 11 | Steps 6 to 10 are repeated for all the elements in $V$ | Z Zernike decomposition corresponding to the aberrations given by all the HP movements in $V$. |
| 12 | The Zernike result matrix is assembled. | Measured Zernike modes matrix $Zmat$ |
| 13 | $Zmat$ is pseudo-inverted | $Z_{HP} = Zmat^+$ |
| 14 | $Z_{HP}$ and all the relevant data are saved. | $Z_{HP}$ file. Data files |
| 15 | Steps 4 to 15 may be iterated increasing the amplitude of $V$. | |

Table 8

### 4.3.2 Result

The procedure in Table 8 was iterated twice: after successfully accomplishing the preliminary flattening procedure (see Sec. 4.2) and after the final flattening procedure (see Sec. 4.6) to improve the process accuracy. The $Z_{HP}$ elements are reported in Table 10 The procedure for the automated optical alignment is presented in Table 9 and has been run, upon request, during the testing activities. The fully automated alignment was requested for long lasting, unattended measurement (see for instance Sec. 4.11); in that case, the procedure was executed once every 15 minutes, which was estimated to be the steadiness time of the optical bench.

**Automatic procedure: HP alignment**

| Step | Action | Result |
|---|---|---|
| 1 | The $Z_{HP}$ matrix is loaded | $Z_{HP}$ matrix |
| 2 | An interferometer image $s$ is taken with $t$ integration time | Image $s$ |
|   | The current Zernike Z decomposition of $s$ is computed | $Z_* = Z\,s$ |
| 3 | The HP correction command is computed with $Z_*$ and $Z_{HP}$ | $HP_* = Z_{HP}\,\#\#\,Z_*$ |
| 4 | The HP is moved with delta command $HP_*$ | HP at new aligned position (within noise). |

Table 9

|        | tip        | tilt       | focus      | comaX      | comaY      |
|--------|-----------:|-----------:|-----------:|-----------:|-----------:|
| **HPx**    | 2.240E+04  | -6.761E+05 | 1.358E+05  | 4.377E+06  | -2.316E+06 |
| **HPy**    | 7.247E+05  | -1.598E+04 | 1.252E+05  | 1.356E+05  | -3.392E+06 |
| **HPz**    | 5.648E+05  | 9.475E+06  | 5.014E+08  | 9.353E+06  | 1.048E+06  |
| **HProtX** | -4.655E+05 | -3.472E+06 | -8.237E+06 | -3.479E+08 | 3.884E+07  |
| **HProtY** | -1.109E+05 | 1.680E+06  | -7.990E+06 | 2.767E+06  | 3.534E+08  |

Table 10 Coupling coefficients between the HP mdegrees of freedom and the Zernike moides measured from the interferometer (HP arbitrary units).

## 4.4 IF collection

The collection of the mirror influence functions (IF) consists in measuring the mirror surface map when a certain (known) deformation is applied. The actuator commands are the eigenmodes of the stiffness matrix, also known as *mirror modes*. The IF collection is the core of the DM characterization, provided that a large part of the following calibration activities is based on the IF data (the flattening computation, the control basis identification, e.g.). Then, it is mandatory to collect the data with a good SNR not to propagate the noise throughout the rest of the process.

The sampling is based on a differential measurement: a sequence of commands is applied on the mirror and captured synchronously by the interferometer; the sequence is composed by a repetitions of mirror modes commands with alternate positive and negative amplitude; we will refer to the repetition pattern of the same mode as the *template*; the length $n$ of the template corresponds to the number of images collected per mode. The synchronization is guaranteed by a frame-to-frame triggering mechanism equipping the interferometer, as described in Sec. 3.1.7. In order to cope with delays in the SW communication between the PC and the interferometer (so that the sequence of commands could start before the frame grabber is completely initialized), a control frame containing a large trefoil deformation is inserted in the sequence of commands. Once the images are collected, they are analyzed according to the template to recover the phasemap given by the mirror shape.

The command amplitude shall be tuned in order to maximize the SNR; two constraints are limiting the maximum applicable amplitude: the fringes density and the requested actuator force. Low-mid order modes amplitude is generally limited by the first point: attention must be paid in verifying that the local deformation, superimposed to the initial mirror deformation (if any) is still within the interferometer capturing range. High order modes amplitude is on the opposite limited by the requested actuator force, while the resulting fringes density is well within range. In this case, the modal amplitude may be automatically reduced by the sampling procedure to match with the specified force threshold. The adopted modal amplitude is shown in Figure 4: the SNR of the measurement may be estimated by comparing the plot with the high order noise as measured in Sec. 4.12.1.

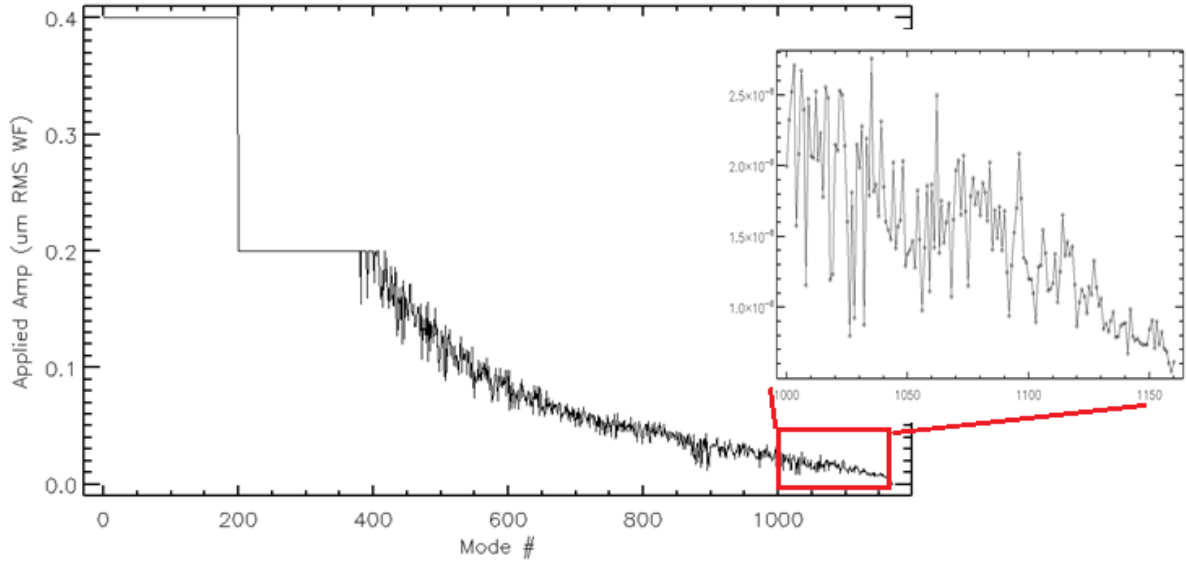

Figure 4 Modal command amplitude for the sampling of the mirror IF.

### 4.4.1 Data sampling

The data collection lasts approximately 5 minutes, when measuring $n=3$ frames per mode. Given such short acquisition time, the system conditions are not changing so that the measurement noise may be considered the same for all the modes. The resulting $n$ images sampled per mode are then analyzed according to the following algorithm to obtain the resulting $S$ optical shape of a given mode:

$$S = \frac{1}{n-1}\sum_{i=2}^{n}\left[(-1)^i \frac{\Phi_i - \Phi_{i-1}}{2}\right] \qquad (1)$$

where $n$ is odd and $\Phi_i$ is the $i$-th image within the sequence with odd and even $i$ with negative and positive modal amplitude, respectively. The algorithm is, essentially, the average of $n$-1 differential images with inverted signs to get rid of the first order drifts. The detailed sampling procedure is provided in Table 11.

**Automatic sampling procedure: IF acquisition**

| Step | Action | Result |
|---|---|---|
| 1 | The relevant parameters are selected: subset of modes, commands amplitude, template length (i.e. number of images to be collected per mode) | List of parameters filled with the chosen value |
| 2 | The sequence of mirror commands is created according to the selected parameters | Sequence of mirror commands: *nacts x n* |
| 3 | The sequence is analyzed to check for actuators exceeding the force threshold. In case, the relative mirror command is scaled to match the force threshold. | Sequence of mirror commands force-limited. |
| 4 | The control frame and the zero commands are inserted in the sequence. | Updated sequence of mirror commands. |
| 5 | A command is sent by the test SW to the interferometer, to start the acquisition. | The frame grabber is initialized, waiting for the first trigger received to capture the first image. |
| 6 | A command is applied and a trigger signal is output by the mirror electronics. | |
| 7 | An image is captured | |
| 8 | Step #6 and #7 are repeated until the end of the sequence | A set of *n* images is captured |
| 9 | The mirror is commanded back to its zero position | |
| 10 | The control frame is searched within the sequence and the first image corresponding to a mirror mode is found. | First good frame in the sequence |
| 11 | The images sequence is analyzed according to the template to compute the wanted phasemap. See below | |
| 12 | Each phasemap is saved together with command amplitude and mode number. | Phasemap file |
| 13 | Metadata are saved | metadata file |

Table 11

**Parameter list: IF sampling**

| Parameter | Value | Description |
|---|---|---|
| $n$ | 3 (also 15 and 51) | number of differential frames in the sampling template |
| $amp$ | 400 nm and 200 nm | Modal command amplitude |
| $m$ | 0-200; 201-400; … | Vector of modes to be sampled |
| $f_{th}$ | 0.6N | Force threshold (maximum actuator force to be applied for the modal command) |
| $c_f$ | 5 | control frame shape, i.e. modal command to be injected in the control frame |

Table 12

### 4.4.2 Result

Many sets of IF have been collected during the test, throughout the iterative process to obtain a flat mirror. Then, the first measurements are limited to a few modes and are needed to build up a (partial) control basis to correct the initial mirror deformations. As soon as the mirror WFE is iteratively lowered, it is possible to sample the high orders: the procedure in Table 11 is therefore iterated starting from mode #0 increasing step by step the number of collected modes.

**Automatic analysis procedure: IF production**

| Step | Action | Result |
|---|---|---|
| 1 | The control frame is found by comparing the RMS of the difference between couple of consecutive images, from the beginning of the sequence. The control frame is the one where the RMS is comparable to that of the control trefoil. | First image valid in the sequence |
| 2 | Starting from the first valid frame, n images are analyzed according to eq (1) to produce the associated phasemap | Final phasemap of the given mode. |
| 3 | The phasemap is saved together with command amplitude and mode number. | File containing the phasemap. |
| 4 | The sequence number of the last processed image is preserved as the starting frame of the next sequence. | |
| 5 | Steps from #2 to #5 are repeated until completion | |
| 6 | Relevant metadata are saved together with the phasemaps. | Files containing FF matrix, command matrix, working actuators, template, are saved. |

Table 13

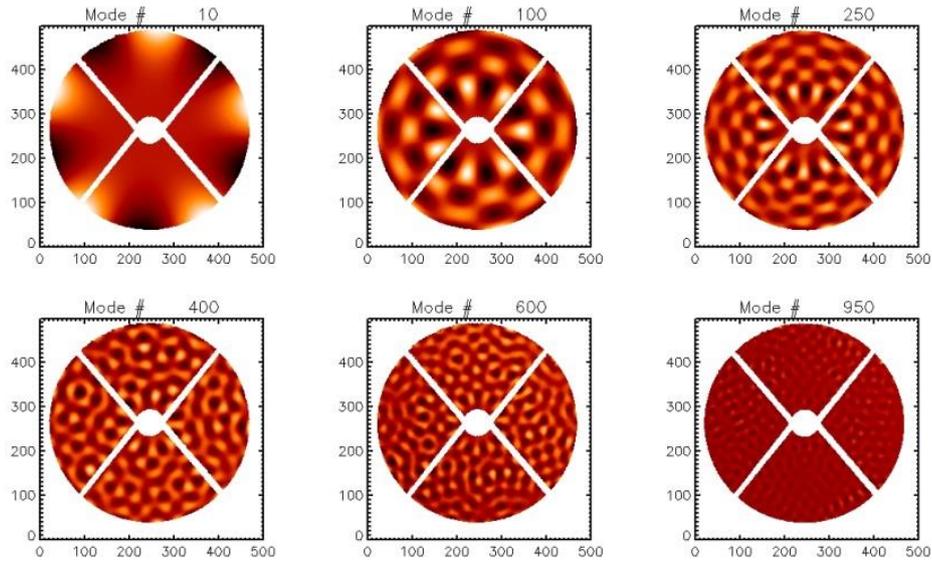

Figure 5 Some mirror modes images, measured with the procedure given in the text.

### 4.4.3 Advanced analysis: IF measurement repeatability

The repeatability of the IF measurements is requested because, especially for high order modes, the SNR of the measurement may be poor, because of the low associated modal amplitude. The repeatability may be evaluated by comparing many optical realizations of the same mode with a reference image; we used as a reference the average image of all the realizations. The comparison is done according to the following procedure.

**Automatic analysis procedure: IF repeatability**

| Step | Action | Result |
|---|---|---|
| 1 | n > 3 IF realizations of IF are selected. | |
| 2 | Starting from the first mode of the IF dataset, the average image of the n realizations is computed. | Average image of the n realizations. |
| 3 | The difference between the average image and each realization is computed. | Differential images wrt average |
| 4 | The tip tilt is removed from each differential frame. | Differential frames, tip tilt removed. |
| 5 | The RMS $r_i$ of each differential frame is computed. The RMS $r_0$ of the average frame is computed. | RMS of differential and average frame. |
| 6 | The ratios $r_i/r_0$ are averaged together and stored as the repeatability error for the given mode. | Repeatability error for the considered mode. |
| 7 | Step from 2 to 6 are executed. | Repeatability errors for the full set of modes. |

Table 14

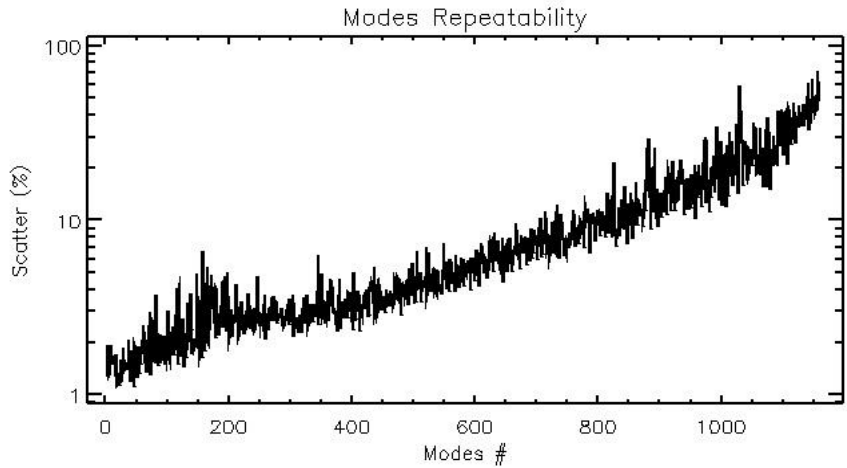

Figure 6 Scatter between a sample of mirror modes images and their average, for the full set of modes.

## 4.5 Interaction and Reconstruction Matrices definition

### 4.5.1 IM definition

The interaction matrix is assembled by piling up all the mirror modes images. It provides the expected mirror shape as seen by the interferometer when a known command is applied: then, it is a projector from the space of the mirror command to the space of the interferometer images. For this reason, it must be insensitive to the mirror position within the image and properly normalized in order to be the most general.

The procedure for the manipulation of the IF images to build up the IM is described below.

**Automatic analysis procedure: IM definition**

| Step | Action | Result |
|---|---|---|
| 1 | The IF datasets to be assembled together are selected. | |
| 2 | The final resolution of the IM is selected. | IM final resolution. |
| 3 | All the masks are loaded and the intersection mask is computed. | IF intersection mask. |
| 4 | The bounding box *b* of the intersection mask is found; the mask is cropped according to *b*. The mask is resampled at the specified resolution. | Cropped, rebinned intersection mask |
| 5 | The index of the valid pixels is saved. | Pixel index. |
| 6 | An empty IM is created, with number-of-modes columns and number-of-pixels rows. | Empty IM |
| 7 | The image corresponding to the first mode is loaded. | |
| 8 | The image is cropped and resampled as the intersection mask. | Cropped, rebinned mode image. |
| 9 | Piston, tip and tilt are removed from the image. The total number of corrected modes (3, in this case) is saved. | Mode image, alignment corrected. |
| 10 | Each pixel of the image is normalized with the command amplitude (metadata saved during the IF sampling). | Normalized mode image |
| 11 | The IM column corresponding with the selected mode number is filled with the valid pixel of the current image. | |
| 12 | Steps 7 to 11 are repeated for each mode | IM filled |
| 13 | IM is saved | IM file |
| 14 | Relevant metadata are saved with the IM: valid pixels index, command amplitude, reform factor, cropping parameters, IF dataset reference, number of corrected modes (tip/tilt and piston, typ.). | Metadata files |

Table 15

### 4.5.2 RM definition

The reconstruction matrix (RM) is obtained by pseudo-inverting the IM (the reason for pseudo-inversion, rather than simple inversion, is that the IM is singular). Then it is a projector from the space of the interferometer image to that of the mirror command. The pseudo-inversion must be performed on valid pixels only within the image. For this reason, when the mirror modes content of an image is to be computed, the most correct procedure is to get the intersection mask of both current image and IM and pseudo-invert it considering the new set of valid pixels. Here, we will describe a general procedure for computing the RM, according to the following steps.

**Automatic analysis procedure: RM definition**

| Step | Action | Result |
|---|---|---|
| 1 | The IM is selected. | |
| | The number $v$ of the corrected modes is loaded (3, typ.) | |
| | The IM is pseudo-inverted (atomic operation), according with the constraints of rejecting $v$ modes. | RM |
| | The eigenmode associated with the least significant eigenvalue (i.e. the lowest) is saved. The | |
| | RM is saved | RM file |
| | Relevant metadata are saved (n, IM reference, eigenvalues) | Metadata file |

Table 16

### 4.5.3 Advanced analysis: cross talk and rejected eigenvectors

**Automatic analysis procedure: IM characterization**

| Step | Action | Result |
|---|---|---|
| 1 | Display the IM intersection mask. Verify that there are no islands in the image. Compute the ratio between the number of valid points and the total pixel within the same circle. | See Figure 7 |
| 2 | Compute the matrices product RM # IM. Compare the diagonal elements with the non-diagonal elements. The non-diagonal elements are given by the modes cross-talk in the data. | See Figure 7 |
| 3 | Plot the RM eigenvalues and verify that there are no points out of the curve …… | See Fig. |
| 4 | Display the image corresponding to the least significant RM eigenvector. Verify that the patterns in the image have a spatial scale smaller than the interactuator one | See Figure 8 |

Table 17

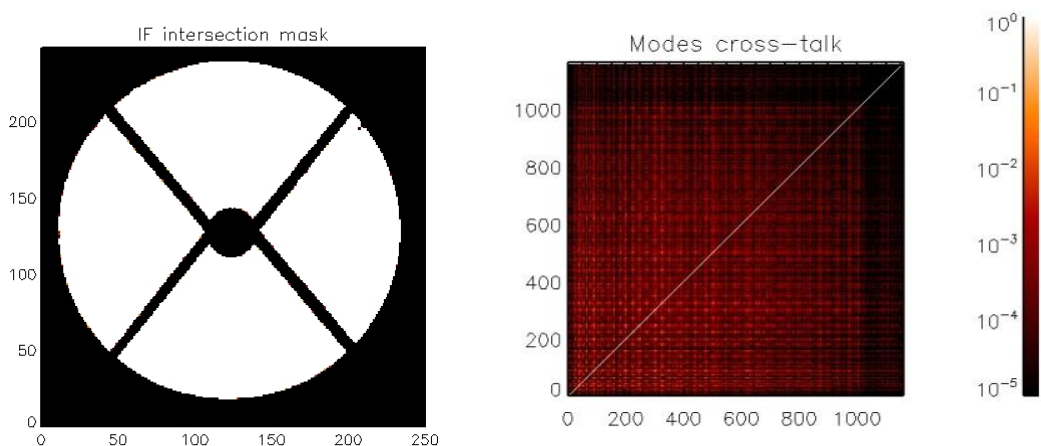

Figure 7 Left panel: mode images intersection mask. Right panel: modes cross-talk matrix.

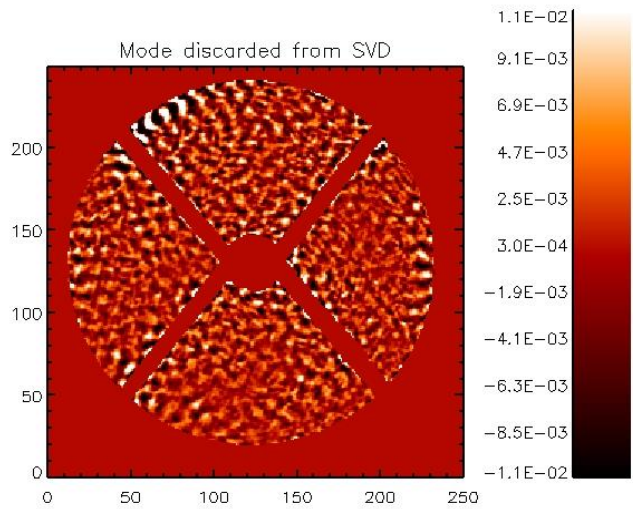

Figure 8 Image of the eigenmode rejected by the singular values decomposition.

## 4.6 Flattening command calibration

The calibration of the flattening command consists in computing the actuator displacement that corrects for a part (or all) of the mirror deformation. The command is composed by a linear combination of mirror modes: the largest the number of modes applied, the better the correction. The mode combination is computed from the current mirror shape provided by the interferometer by projection to the modal basis: therefore, the largest number of modes that can be applied is given by the dimension of the modal basis. As discussed in Sec. 0, for the sampling of the modal basis the fringes density on the mirror must be such as not to exceed the interferometer capture range, when the modal command is applied. Then, both IF sampling and mirror flattening are iterative procedure: as soon as the mirror is corrected with $n1$ modes, a new set of $n2>n1$ modes is sampled and subsequently corrected and so on.

The accuracy of the flattening result and of its verification is limited by the measurement noise; here, we are not considering the intrinsic degeneracy of the alignment modes (focus and coma). The most limiting noise is air convection, whose contribution may be evaluated with the procedures in Sec. 4.12. Here, we will just underline that the reference and the result images will be integrated over a given time span (minutes) in order to average out the convection.

### 4.6.1 Data sampling

The procedure for the calibration of a flattening command is depicted below.

**Automatic sampling procedure: Flattening calibration and measurement**

| Step | Action | Result |
|---|---|---|
| 1 | The modal basis or (optionally) the subset of modes to be used are loaded together with relevant metadata (including also image sampling parameters and mirror command matrix CM). | Modal basis |
| 2 | The alignment modes *A* not to be corrected are defined (piston, tip-tilt, power and coma, typ.) | |
| 3 | | |
| 4 | An interferogram image *S* is taken (averaged over a given time span), together with its mask. | Reference mirror image and associated mask. |
| 5 | The mask is cropped according to its bounding box and resampled with the same sampling parameters of the modal basis images. | Cropped and resampled image mask. |
| 6 | The intersection between the image mask and the IF mask is computed and the valid pixel index saved. | Intersection mask |

| 7 | The modal basis images are masked according to the new intersection mask. | Masked modal basis. |
|---|---|---|
| 8 | The tip-tilt is removed from all the modal basis images. | modal basis, images re-aligned. |
| 9 | The RM is computed (with $A$ discarded modes). | RM |
| 10 | S is matrix-multiplied with RM to obtain the modal coefficients $M$. | $M$, corresponding to the mirror modes decomposition of the reference image. |
| 11 | The actuator command $C$ is obtained by multiplying the modal coefficient $M$ with the command matrix CM. Its sign is changed. | Actuator command $C$ |
| 12 | C is applied to the actuators. | |
| 13 | The result image S1 is taken (averaged over a given time span), together with its mask. | Flattening result image. See Figure 10. |
| 14 | The flattening command, the reference and result image, RM and IM references are saved. | All relevant data saved. |

Table 18

**Parameter list: Flattening calibration and measurement**

| Parameter | Value | Description |
|---|---|---|
| $m$ | up to 1170 | number of the control matrix modes to be corrected |
| $p$ | 90 | total number of interferometer frames to be averaged |
| $t$ | 10s | time delay between two interferometer frames |
| $z$ | 6 | number of Zernike modes to be removed from the reference image |

Table 19

### 4.6.2 Result

Within the optical test, we considered the starting point of the flattening procedure the time when the optical pupil was fully visible. The mirror shape at this stage is shown in the left panel of Figure 9, where the whole pupil is visible even if many spots have a large fringes density.

The result we present here (in Figure 9, right panel) is referred to the last iteration of the procedure, when all the mirror modes are applied.

In Figure 10 we show the resulting image *S1*, as described in the step # 13 of Table 18. The image has been integrated over a time span of 15 minutes, averaging together 90 interferometer frames. The WFE of the image is 30 nm RMS, after removing the alignment aberrations.

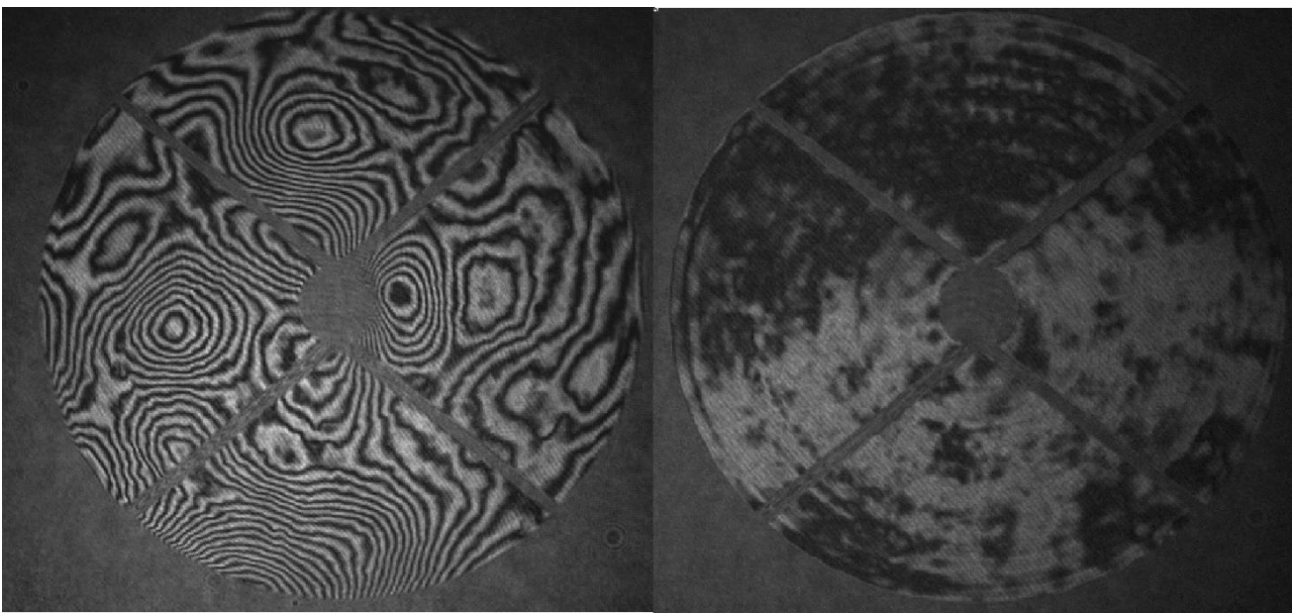

Figure 9 Left panel: mirror interferogram at the beginning of the flattening procedure. Right panel: mirror interferogram after the last step of the flattening procedure.

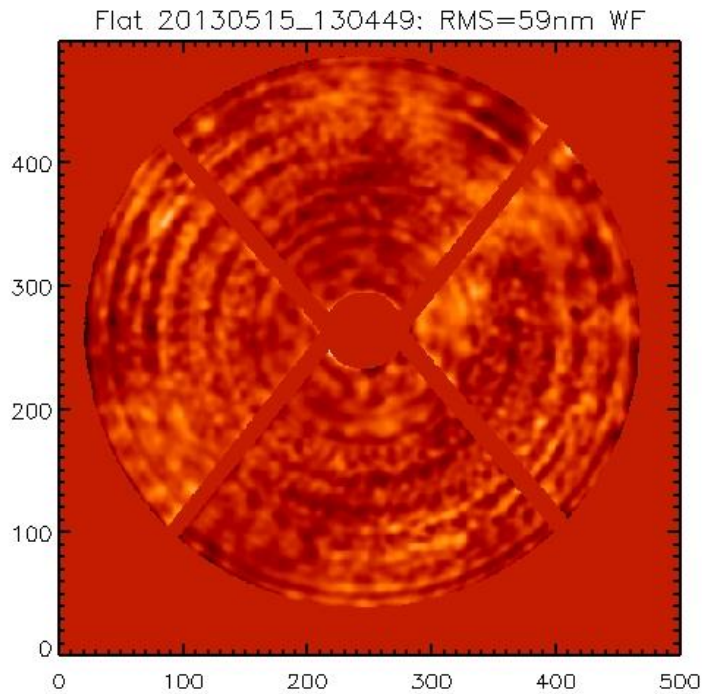

Figure 10 Resulting image of the flattening procedure, after subtracting the 6 alignment Zernike modes.

#### 4.6.3 Advanced analysis

The advanced analysis after the flattening procedure is a sanity check of the applied correction.
The first stage is devoted to quantify the degree of correction, by simply computing the residual WFE as the RMS of the pixel value within the pupil: as the interferometer image *S1* is blended at any time by alignment drift of the optical bench, the alignment aberration are removed before the computation. Additionally, the first 10 or 15 Zernike terms may be also subtracted to take into account the low order thermo-mechanical deformations of the supporting structures induced by drifts during the *S0* and *S1* sampling time span. These two numbers are a general qualification of the result of the flattening procedure; their variation throughout the iterative process is monitored to check the progress in flattening the mirror. The result is in the end compared with the requirements.

The second stage of the advanced analysis is devoted to the analytic estimation of the flattening result and its comparison with the actual *S1* shape. The computation is performed by adding to the *S0* shape the actuator flattening command multiplied by the IM: the operation consists in creating the optical shape of the corrected mirror, skipping the second measurement *S1* (which is affected by measurement noise).

A third analysis stage is the computation of the modal content of the *S1* image (by projection on the RM) to verify that no modes have been excited by the flattening process.

**Automatic analysis procedure: Flattening residuals**

| Step | Action | Result |
|---|---|---|
| 1 | Compute the residual WFE RMS after removing the alignment aberrations (piston, tip-tilt, power and coma). | WFE after alignment |
| 2 | Compute the residual WFE RMS after additionally removing the low order aberrations coming from thermo-mechanical drifts (astigmatism and trefoil). | WFE after removing 10Zernike modes, corresponding to alignment and mechanical deformations. |
|  | Compute the synthetic flattening result $S2_*=S1-IM*f$, where $f$ is the actuator command applied. | Synthetic flat $S2_*$. See Figure 11, left panel |
| 3 | Compute the modal content *M1* of the flattening residual by multiplying the image *S1* with the RM. | Modal residual. See Figure 11, right panel. |

Table 20

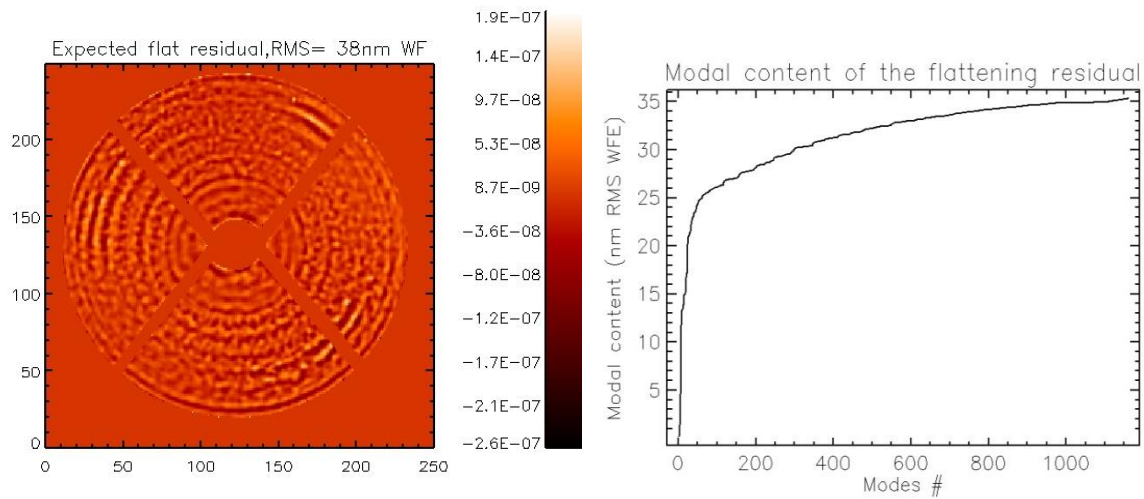

Figure 11 Left panel: expected flattening result; right panel: modal content of the measured flattening result.

## 4.7 Actuator mapping

The actuator mapping consists in identifying the location corresponding to the capacitive sensor area within an image provided by the interferometer. Such information is requested to find the actuator associated with a given image region (for instance, when a bump is observed) or more generally to correlate the position value provided by the capacitive sensor with that given by the interferometer signal. Such task is for instance at the base of the capacitive sensors calibration procedure.

The routine is based on the assumption that the actuator location on the image corresponds to the center of its influence function. Such idea, however, fits poorly with those actuators that are poorly visible (for instance, those obscured by spider arms or central obstruction) and those on the outer edge where, given the lack of external constraints, the peak of the influence function is located at the glass edge.

The mapping is obtained by fitting the as-built actuator coordinates with the measured one, starting from the IF data; the IF of the actuators not matching the criterion given above are excluded from the fitting; the model is based on a rigid body transformation.

The output of the procedure is a dataset containing the center and the radius of the IF images used, the location of the actuators center on the image and the image scale. These dataset is sufficient to identify the actuator locations on any image, irrespective of its size and mirror position within it.

A by-product of the procedure is the sign calibration of the interferometer signal.

**Automatic analysis procedure: Actuator mapping**

| Step | Action | Result |
|---|---|---|
| 1 | A full set of zonal IF is selected. | IF dataset |
| 2 | A reference image is loaded; the center and radius of the mask are found. | Image |
| 3 | Mask center and radius and image size are saved in the destination folder. | Mask parameters |
| 4 | the actuator coordinates are restored from a file | Actuator coordinates by desing |
| 5 | the list of useful actuator for mapping is built, by excluding those whose IF are poorly imaged | Actuator list |
| 6 | The peaks of the IF are located. | IF peak coordinates |
| 7 | The sign of the peak is noted and compared with the sign of the given actuator command. | Peak sign |
| 8 | The IF with a peak value below a given threshold are excluded (as they correspond to obscured actuators). | |

| 9 | The set of coordinates obtained is fitted with the as-built ones to identify the rotation and magnification of the image with respect to the pattern designed. | Geometry fitting parameters |
|---|---|---|
| 10 | The fitting residuals are analyzed to test the hypothesis. | |
| 11 | The coordinates of all the actuators are computed upon the model parameters obtained. | Actuator coordinates in the image |
| 12 | The pixel indexes associated with each capacitive sensor area are computed. | Capacitive sensor areas in the image |

Table 21

### 4.7.1 Result

The result of the mapping is shown in Figure 12: in the left hand panel, the reference image is shown, with the identified actuator locations highlighted. In the right hand panel, a zoom is given, showing the actuator IF images (3, summed together); the actuator locations is highlighted to show the matching.

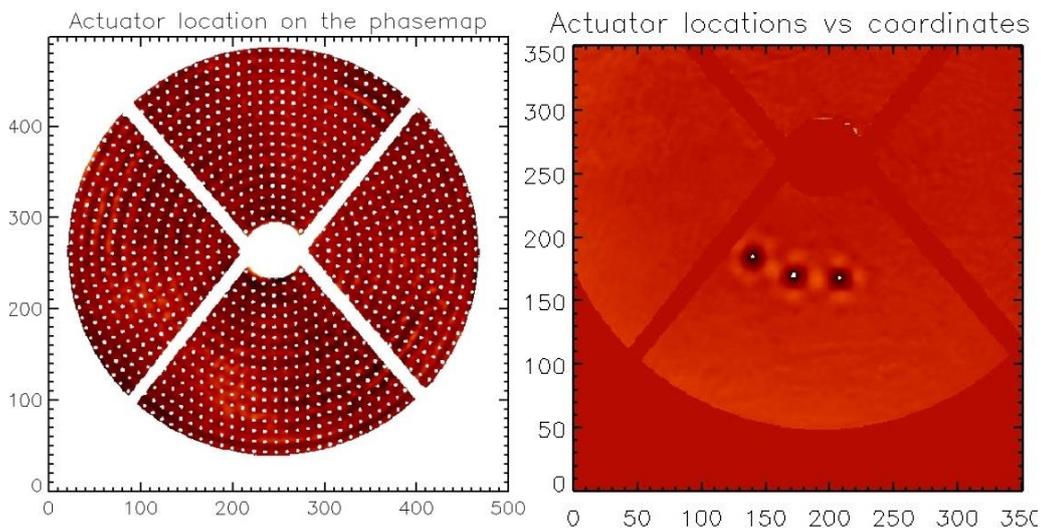

Figure 12

### 4.7.2 Advanced analysis

The intrinsic limitations of the procedure are the images resolution and the accuracy of the actuators coordinates data provided. Such limitations affect the mapping accuracy, inducing a systematic error when measuring the interferometer signal associated with a given actuator position: such error is dependent on the image to be processed.

We evaluated the optical measurement error for two specific cases: the capacitive sensor calibration images and the flattening result images. The analysis procedure it is based on a rigid offset of the actuator coordinates on the image, according to a square grid with convenient size; we tested both 3 and 5 pixels, corresponding to the interactuator distance (i.e. with such mapping error, any actuator will be measured at the location of the closest neighboring one). The actuator signal from the interferometer was then measured drawing a capacitive sensor area centered on each pixel of the grid. The full procedure is described in Table 22.

**Automatic analysis procedure: actuator optical measurement error.**

| Step | Action | Result |
|---|---|---|
| 1 | A reference image is selected | Image |
| 2 | A maximum mapping error is chosen | 5x5 pixel grid |
| 3 | The actuator mapping file is loaded | actuator coordinates |
| 4 | The location of the first actuator on the image is computed; the pixel indexes of the corresponding capacitive sensor are found. | pixel indexes, associated with a capacitive sensor area |
| 5 | the mean value of the image is computed over the selected indexes. | optical position value associated with the capacitive sensor. |

| 6 | the actuator location is moved to each of the points in the grid | |
|---|---|---|
| 7 | steps 4 and 5 are repeated | |
| 8 | A map of the resulting values is created. | optical positions associated with a mapping error on the image. |

<div align="center">Table 22</div>

## 4.8 Capsens calibration

The interaction matrix mechanism allows obtaining a local calibration of system, i.e. to know the optical response when a small command is injected starting from the nominal working position. No information is obtained about the system behavior far from the standard gap. We are also limited by the interferometer capturing range (of the order of 10 um), so that we cannot sample very large command to obtain a greater range calibration, and by the fact that an interferometer is not sensitive to the piston signal.

To be able to perform large commands the system must be calibrated in a different fashion. Here we will describe an absolute, large stroke calibration of the capacitive sensors: the procedure consists in measuring the Z position (with respect to the interferometer) of all the actuators while they are moved across their working range.

The main concept behind the procedure is that an uncalibrated actuator performs a null command with very high precision, as its position sensor noise is as low as 2~5 nm RMS. Then, if the mirror is moved between two positions having different piston value and one or more fixed actuators, it is possible to resolve the piston uncertainty with the a priori knowledge of the fixed actuators position (markers).

We selected two sets of 3 markers each, as in Figure 13, in order to define a plane accordingly and eliminate automatically the tip-tilt noise induced by vibrations. The mirror is then moved with trefoil commands that maximize the deformation having three fixed points; to further reduce measurement noise, each command is applied and sampled with the push-pull technique (as in 4.2): the reading of the capacitive sensors is saved at each step; in the end, the resulting images are piled up together according to the marker positions and the Z measurement of each actuator is compared with the associated optical one.

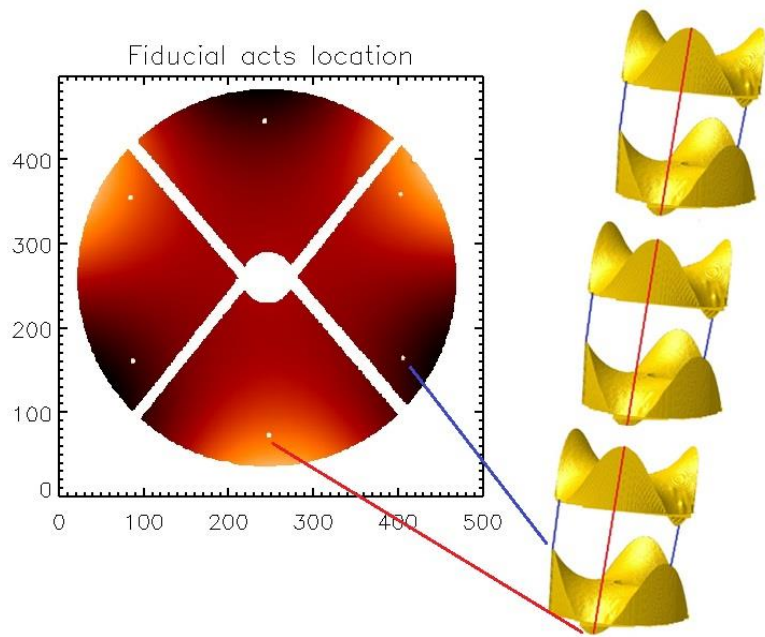

Figure 13 Location of the marker actuators

### 4.8.1 Data sampling

**Automatic sampling procedure: Capacitive sensors calibration**

| Step | Action | Result |
|---|---|---|

| 1 | Three marker actuators $a1$ are identified: they should be placed at the second/third outermost ring of actuator, far from not working actuators, and at 120° relative position. Three other marker actuator $a2$ are identified, with the same constraint as for $a1$ and located at a 60° angle with respect to them. | Marker actuators $a1$ and $a2$ |
|---|---|---|
| 2 | A command template $T$ is created by considering the trefoil mirror mode. | |
| 3 | The value of $T$ on the markers $a1_1$ and $a2_1$ (first actuator within each set) is registered. | $T(a1_1)$; $T(a2_1)$ |
| 4 | A command $C$ is created with a linear combination of the first 6 mirror modes:<br>$C(a1)=T(a1_1)$;  $C(a1)=T(a1_1)$;<br>$C(a2)=T(a2_1)$;  $C(a2)=T(a2_1)$; | Command C that is levelled on $a1$ and $a2$ markers. |
| 5 | The first mirror command is defined: $P_1=C- C(a1)$; (zero displacement on $a1$). It is scaled to match the desired stroke $v$ | Mirror command $P_1$ |
| 6 | The next mirror command is created: $P_2= C+P_1(a2)$, i.e. zero displacement on $a2$.<br>The generic command is created as<br>$P_x=C+ P_{x-1}(a1)$;<br>$P_{x+1}=C+ P_x(a2)$<br>Commands are created to span the calibration range $[-z1; +z2]$ (both positive and negative movements, with respect to the nominal position). | Mirror command $P_2$; generic mirror command $P_x$ |
| 7 | Each command $P_x$ is replicated in a sequence $S_{Px}$ of $n$ odd positive and negative iterations, according to the template $[P_x,-P_x,P_x...]$ | $S_{Px} = [P_x,-P_x,P_x, (-1)^{n+1}P_x]$ |
| 8 | Each sequence of commands $S_{Px}$ is applied to the mirror and sampled with the synchronization mechanism described in Sec. 4.4 | Dataset of n differential frames, one dataset for each of the $S_{Px}$ sequences. |
| 9 | The capacitive sensor values are read for each step. | Capacitive sensor values. |
| 10 | the images and capacitive sensor readings are saved | Data files |
| 11 | The relevant metadata are saved | Metadata files |

Table 23

**Automatic analysis procedure: Capacitive sensors calibration**

| Step | Action | Result |
|---|---|---|
| 1 | The dataset is restored | Set of images |
| 2 | The actuator mapping is restored | Actuator mapping |
| 3 | The markers position in the image is computed. | Pixel indexes of marker actuators |
| 4 | The images dataset is analyzed with the differential algorithm in Sec. 4.4.1. | Dataset of resulting differential images, 1 frame per command |
| 5 | An empty frame $Z_0$ is created | Null $Z_0$ frame |
| 6 | The differential image $S_i$ is considered | |
| 7 | The value of the image on the non-fixed markers is measured | Pixel values on markers |
| 8 | The plane $P_i$ passing by the three points is computed | Plane $P_1$ |
| 9 | The differential image $S_{i+1}$ is considered; the absolute image $Z_{i+1}$ is computed as $Z_i+S_{i+1}+P_i$ | Absolute $Z_{i+1}$ image |
| 10 | Steps 6 to 9 are repeated for all the differential images; in the end, the sequence of $Z$ absolute images is obtained | Sequence of $Z$ images |
| 11 | The pixel values of any actuator is measured on $Z_i$ image | optical absolute position of all the actuators |
| 12 | The capacitive sensor readings are fitted versus the optical reading; the calibration coefficients of all the actuators are computed | Calibration coefficients. |
| 13 | The full procedure is iterated enlarging the calibrated span. | Updated calibration coefficients. |

Table 24

**Parameter list: capacitive sensor calibration**

| Parameter | Value | Description |
|---|---|---|
| *a1* | 692, 661,723 | first set of marker actuators |
| *a2* | 646, 677, 707 | second set of marker actuators |
| *v* | 1.5um | command stroke (difference of piston value between two consecutive commands) |
| *T* | 5 | Trefoil template (mirror mode number) |
| *z1* | -5um; -35um | negative maximum distance |
| *z2* | +5um; +40um | positive maximum distance |

Table 25

### 4.8.2 Result

The output of the procedure is the set of the calibration coefficients, between the capacitive sensor reading and the optical absolute position as computed in Table 25. The full procedure is iterated as long as the calibrated system is able to span a larger movement range. Running the procedure with a preliminary calibration results in a gap between optical and electrical positions; this gap is larger when the system is moved far from the nominal working position. As long as the calibration is iterated, the gap is compensated: it may be taken as a quality indicator of convergence. In Figure 14 two typical plots are shown. In Figure 15 the optical versus electrical reading is given.

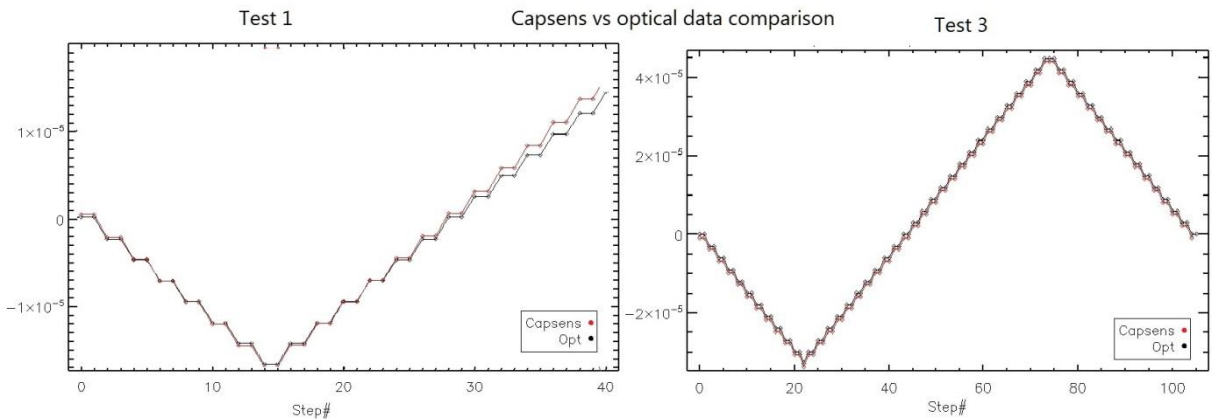

Figure 14 Left panel: electrical and optical position, actuator #791, measured at the first iteration of the procedure. The departure between the two dataset is visible on the right. Right panel: electrical and optical position, actuator #791, measured at the 4$^{th}$ iteration of the calibration procedure.

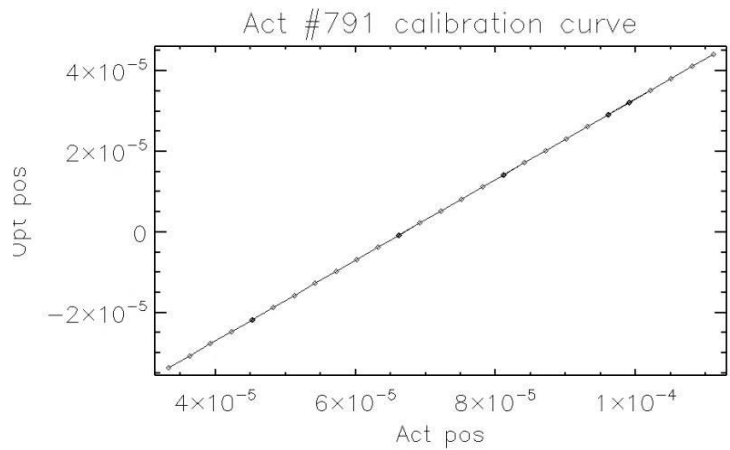

Figure 15 Electrical versus optical position, actuator #791, measured at the 4$^{th}$ iteration of the calibration procedure.

### 4.8.3 Advanced analysis

The procedure is based on the use of fixed actuators as markers to join the images and reconstruct a piston movement. The marker must produce a zero movement with a zero command; any deviation from such hypothesis will inject noise in the calibration process. This assumption may be tested by comparing two capacitive sensor readings when an identical command is given, for instance during the sequence of positive and negative $P_x$ commands. An example image is given in the left panel of Figure 16; the result (for the markers) is given in the right panel, and for all the actuators in the upper frame of Figure 17.

The calibration coefficients, measured twice at the same stage of the process, may be plotted together to estimate the repeatability of the measurement. The result is given in Figure 18.

**Automatic analysis procedure: Capacitive sensors calibration**

| Step | Action | Result |
|---|---|---|
| 1 | The capacitive sensor readings are restored | actuator position from capacitive sensors |
| 2 | The standard deviation $\sigma_{ai}$ of the values associated at the same command is computed | standard deviation of |
| 3 | Step 2 is repeated for each actuator $a$ | |
| 4 | steps 2 and 3 are repeated for each frame $i$ within the dataset | |
| 5 | The $\sigma_{ai}$ values are shown as an image | See Figure 17. |
| 6 | The $\sigma_{ai}$ values referred to the markers only are plotted. | See Figure 16, right panel. |
| 7 | The calibration coeffiecients (computed as in step 12 in Table 24) associated with two or more measurement at the same stage are plotted together. | See Figure 18 |

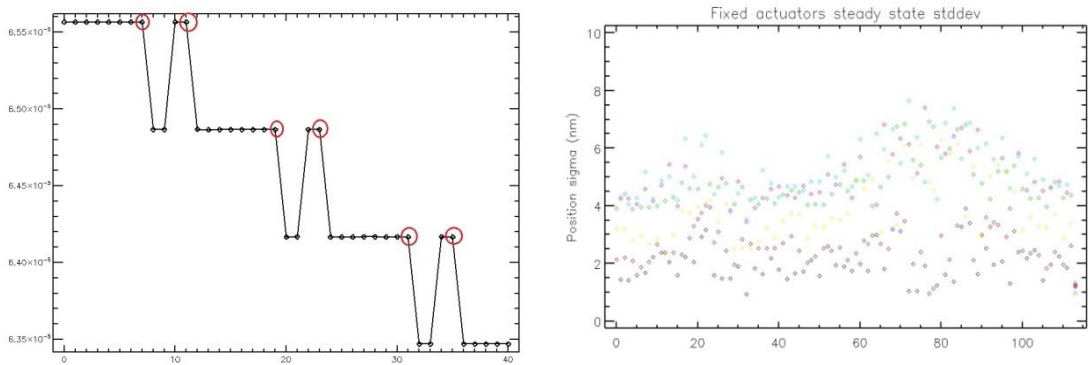

Figure 16 Left panel: actuator command (the dataset is referred to a marker) during the procedure, to highlight the

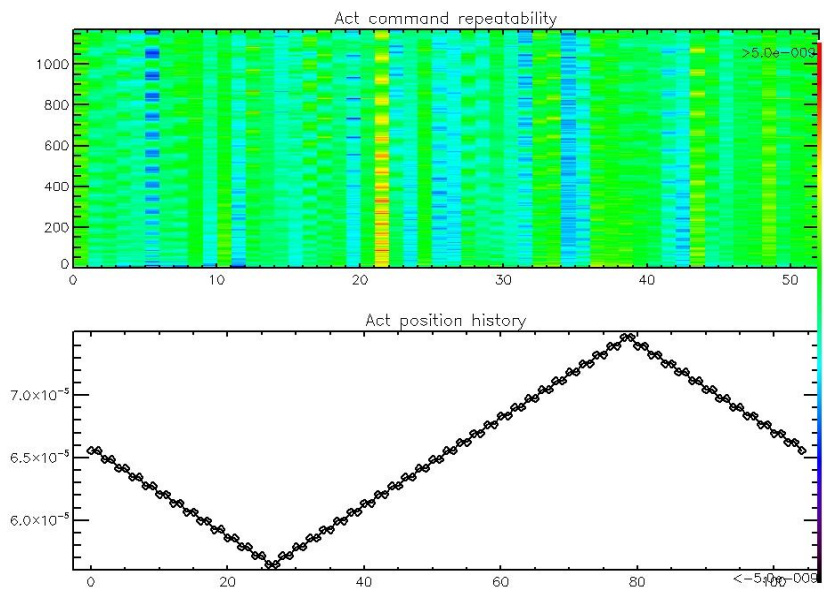

Figure 17 Top panel: color scale representation of the actuator position standard deviation during the command history; horizontal axis: command history step; vertical axis: actuator number. Bottom panel: command history (position) of a marker actuator.

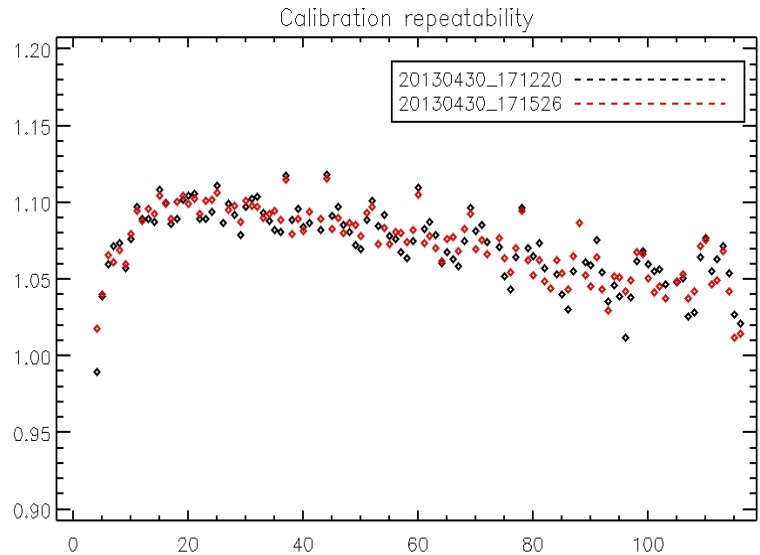

Figure 18 Capacitive sensor calibration coefficient (sample of 120 values) as measured during two separated tests, to show the repeatability of the measurement.

## 4.9 Field stabilization test

The field stabilization is a pure tilt command which is requested to compensate for the telescope vibrations. The command is computed by projecting the geometrical tilt to the mirror modes; in order to keep the requested forces low, only low order modes are summed up. For the verification of the field stabilization command, we are interested in: command precision (i.e. its departure from the geometrical tilt), its repeatability, its accuracy (i.e. departure of the applied tilt from the desired value).
Given that the requested stroke may be larger than the interferometer capture range, the test was divided into two sessions, a small stroke and a large stroke one.

### 4.9.1 Small stroke

The verification has been performed by applying a sequence of equal amplitude, opposite sign field stabilization commands. The sequence was applied and sampled at high frequency (20Hz); the data collection was synchronized with the command application by mean of the trigger line. The command amplitude was 2", corresponding to a mechanical displacement of 10 um PtV measured on the shell edges: the associated fringes density is within the interferometer capture range, thus allowing a direct measurement. Both X and Y axis tilt have been measured.
The sampling procedure is given in Table 26.

*4.9.1.1 Data sampling*

**Automatic sampling procedure: Small stroke field stabilization test**

| Step | Action | Result |
|---|---|---|
| 1 | The RM is loaded with its mask. A mode threshold $i_{max}$ and a field stabilization command amplitude $a$ for the test are selected. | RM |
| 2 | A geometrical tilt $t$ is created on the IM mask, with unitary amplitude. | Geometrical tilt on the mask |
| 3 | The modal decomposition of $t$ is computed by projection on the RM: $v$ = RM ## $t$. | Modal decomposition of the geometrical tilt. |
| 4 | v is filtered according to $i_{max.}$ | Modal tilt, low order modes only. |
| 5 | A sequence of $n$ commands [$av, -av, av, -av$] is created. | Sequence of opposite tilt commands |
| 6 | The DSM is put in SET mode and the interferometer is aligned. | |

| 7 | The full sequence of command is applied and sampled with the trigger synchronization mechanism. | Dataset |
|---|---|---|
| 8 | The dataset is saved. | Data files |

Table 26

**Parameter list: small stroke field stabilization test**

| Parameter | Value | Description |
|---|---|---|
| $n$ | 100 | number of differential frames |
| *amp* | 2" | tilt amplitude |
| $i_{max}$ | 30 | maximum mode number to fit the geometrical tilt |

Table 27

The dataset was analyzed with a differential algorithm, i.e. by computing the differences between consecutive images, and summing then together the resulting frames. By mean of the differential analysis, the slow alignment drifts (tilt) and convection are rejected; also, as the alignment is not modified throughout the sampling, it is possible to measure the absolute tilt movement performed (i.e. removing the tilt uncertainty of the optical system).

#### 4.9.1.2 Result

The dataset is analyzed to measure the resulting tilt and the high order deformations, as in the procedure in Table 28. The results are shown in Figure 19 and Figure 20.

**Automatic analysis procedure: Small stroke field stabilization test**

| Step | Action | Result |
|---|---|---|
| 1 | The dataset is loaded | Dataset, $n$ images |
| 2 | A $2i$ image within the sequence is subtracted to the previous one. | Differential image $\Delta_i$. |
| 3 | Tip and tilt of $\Delta$ are measured and stored. | $i^{th}$ command tip and tilt values $tt_i$ |
| 4 | The tip-tilt angle $\gamma_i = arctg(tip/tilt)$ is computed. | Tip-tilt angle $\gamma_i$. |
| 5 | The alignment aberrations are subtracted from $\Delta$ and the WFE is computed. | $i^{th}$ command WFE. |
| 6 | Steps 2 to 5 are repeated for any $i<n/2$. | |
| 7 | The $n/2$ differential frames, after alignment, are averaged together. | Averaged differential image. |
| 8 | The WFE of the average frame is computed. | Average frame WFE. |
| 9 | $\gamma$ is plotted. | |
| 10 | The mean value and standard deviation of $\gamma$ is computed. | Average and standard deviation of $\gamma$ sample. |
| 11 | $tt$ is plotted. | |
| 12 | The mean value and standard deviation of $tt$ is computed | Average and standard deviation of $tt$ sample. |

Table 28

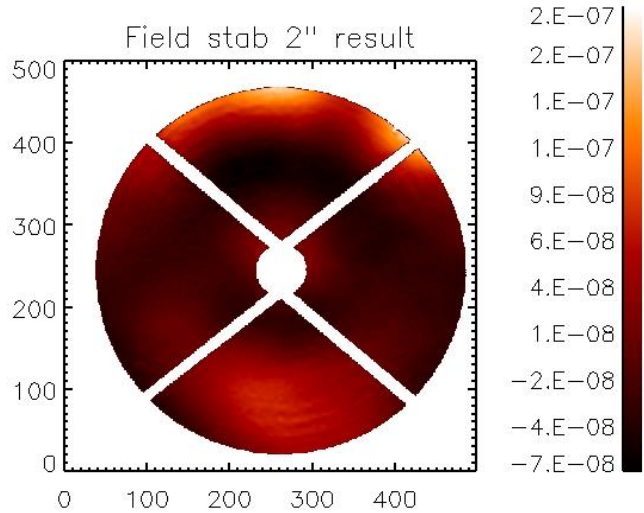

Figure 19 Field stabilization result, DSM shape at 2"tilted position (tilt removed). RMS =81 nm WF

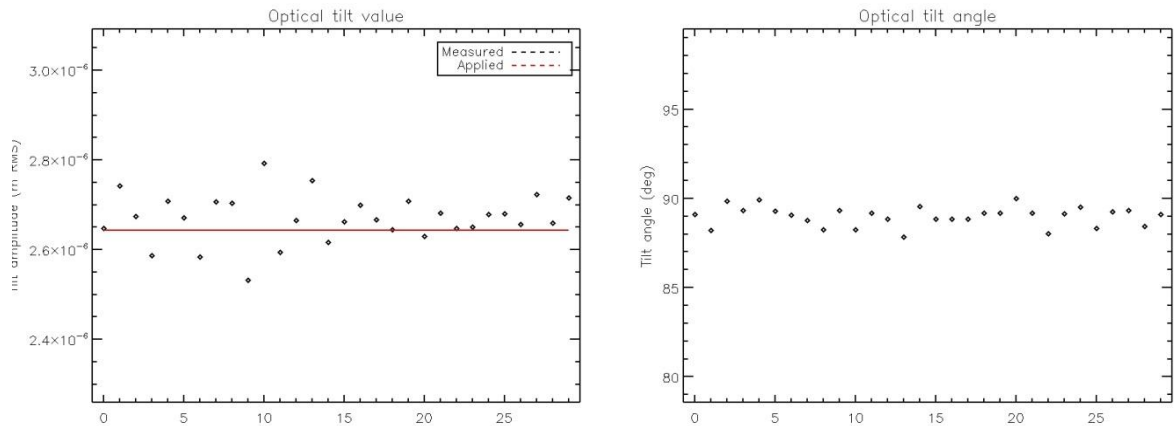

Figure 20 Accuracy of the tilt command in field stabilization mode. Left panel: amplitude of the optical tilt, compared with the applied command. Right panel: tip-tilt orientation (the absolute value is depending from the DSM and interferometer orientation), showing a peak to valley spread of 2°.

#### 4.9.2 Large stroke

A second test is devoted to the verification of the field stabilization command in the large stroke regime; here, the commanded tilt is 12", corresponding to 60um PtV displacement measured on the shell edge. The fringes density associated with such large stroke is out of the interferometer capturing range, so that it is not possible to measure it directly. The procedure is based on a sequence of smaller, incremental tilt movements, followed by a HP movement to restore the optical alignment. An image is taken after each step; as the sampling method is an absolute measurement, time integration is requested to average the convection noise out. A reference image will be sampled with the DSM at its nominal position to subtract the constant offset like the polishing and flattening residuals.

As the final position is achieved composing together two movements (both DSM and HP tilt) it is not possible to measure the tilt accuracy of the absolute command.

##### 4.9.2.1 Data sampling

The dataset has been collected according the procedure in Table 29; each image has been sampled by averaging 50 frames collected over 5 minutes. The full procedure lasted 1 hour per tilt direction.

**Automatic sampling procedure: large stroke field stabilization test**

| Step | Action | Result |
|------|--------|--------|

| 1 | The DSM is put in SET mode and the HP is aligned. An image is taken by averaging together *n* frames over a time span *t*. The image is saved. | Image and data file associated with the the DSM nominal position. |
|---|---|---|
|   | A sequence *V* of tilt amplitude is selected, provided that each individual differential movement produces an optical tilt visible by the interferometer. | Tilt amplitudes *V* to be applied. |
| 2 | The unitary tilt command is computed as in Table 26, steps 2 to 4 and scaled with *V*. | Sequence *T* of tilt commands to be applied |
| 3 | The first command $T_0$ is applied | DSM at $T_0$ position |
| 4 | The system is aligned by the HP |   |
| 5 | An image is taken by averaging together *n* frames over a time span *t*. The image is saved. | Image and data file associated with the $T_0$ position. |
| 6 | Steps 3 to 5 are executed for all the commands in *V*. | Sequence of images associated with any *T* positions. |
| 7 | Steps 3 and 4 are executed with reversed order (e.g.: 12", 9", 6",3",0") | DSM and HP at their nominal positions, no images collected. |
| 8 | Steps 3 to 7 are repeated with negative amplitude (-*T*) | Sequence of images associated with any *T* positions. DSM and HP back at their nominal positions. |
| 9 | The command sequence and the relevant metadata are saved with the proper tracking number. | Metadata file. |

Table 29

**Parameter list: large stroke field stabilization test**

| Parameter | Value | Description |
|---|---|---|
| *V* | 3"; 6";9";12 | Sequence of tilt commands |
| *n* | 30 | number of frames to be averaged together |
| *t* | 300s | total integration time to produce a single image |

Table 30

*4.9.2.2 Result*

The dataset processed as in Table 31 is shown in Figure 21, were *ap, bp* and *d* are given. After the correction for the flattening high order residuals (subtracting f0 and f1), we measured a 133 nm RMS and 124 nm RMS offset from a pure tilt (which was removed from the images), respectively when commanding a +12" or a -12" position.

The DSM precision may be estimated from the *d* image, showing a WFE of 25 nm RMS, a value that is fully compatible with the measurement noise (see also Sec. 4.12.2). It follows that two opposite commands, spanning the full mirror working range, produce two opposite shapes whose sum is zero within noise. The bumps in the *d* image are produced by two specific actuators having noisy calibrated internal metrology.

According to this data, it is not possible to disentangle the effect of any offset in the field stabilization command computation and that of a global mirror calibration error (accuracy error); they contribute jointly to the measured 130 nm RMS departure of the achieved shape from the geometrical tilt. As the DSM position precision has been demonstrated, it is possible to tune the field stabilization command in order to obtain an accurate geometrical tilt result.

**Automatic analysis procedure: large stroke field stabilization test**

| Step | Action | Result |
|---|---|---|
| 1 | The dataset is opened. | Sequence of images, arranged according to applied command (e.g.: 0",3",6",9",12",0",-3"…) |
| 2 | The images associated with farthest points and zero are considered. We will indicate them as: [$f_0,a,f_1,b,f_2$], where *f* stands for *flat*. | Dataset of flat DSM images and 12" images. |
| 3 | The 12" images are processed as: $ap = a-(f_0+f_1)/2$ | Processed 12" image. |
| 4 | *ap* is corrected by removing alignment aberrations | 12" image, aligned. |
| 5 | The WFE of *ap* is computed | Residual WFE at 12". |
| 6 | Steps 3 to 5 are repeated for *b*. | -12" image and WFE. |

| 7 | The noise of the procedure is evaluated by computing $d=(ap+bp)/2$. | Measurement noise |

Table 31

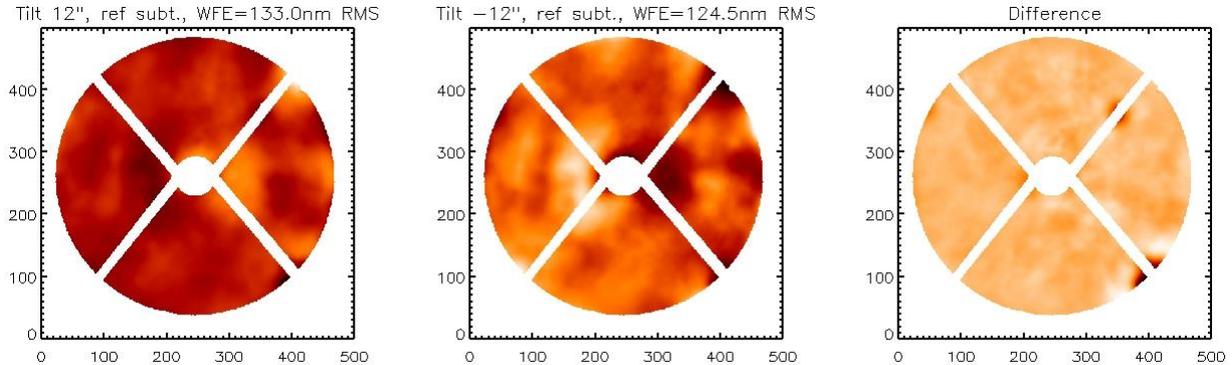

Figure 21 Field stabilization, large stroke test results. Left and middle panel: differential shape corresponding to the 12" and -12" tilt movements; right panel: difference between the two opposite shapes.

## 4.10 Optical step responses

The step response test is devoted to the measurement of the temporal mirror response to a step signal, to estimate the system settling time, overshoot and steady state error. Such test is part of the electromechanical characterization: in particular the test is requested to verify that the mirror bandwidth is compliant with the desired AO loop frequency (1kHz, e.g.).
The optical verification is challenging, provided that a sub-millisecond temporal resolution is requested while commercial frame grabbers frame rate are not faster than 100Hz. To overcome this issue, we implemented a low frequency, phase locked sampling procedure, described in Table 32. Such procedure may be applied for the optical high frequency measurement of any mirror command (within the interferometer capture range).

### 4.10.1 Data sampling

The measurement is based on the a priori knowledge of the temporal delay between the command application and the sampling time. We recall here that the system is equipped with a synchronization mechanism (see Sec. 3.1.7): when a command is applied to the mirror, a TTL signal is output to the interferometer to grab a single image. The delay between the detection of the TTL edge and the exposure is configurable by the interferometer software. A collection of images captured at different delays may be rearranged as a sequence of high frequency frames. The number of datasets to be collected depends on the time span to be measured and by the desired sampling frequency: for instance, to collect data in the time span 0 to 2ms since the command application, at a final sampling rate of 10kHz, 21 datasets are requested.
The basic sampling mechanism is the same for the IF measurements (refer to Sec. 4.4).

| **Automatic sampling procedure: Optical step responses** | | |
|---|---|---|
| **Step** | **Action** | **Result** |
| 1 | A set of command $C$ to be sampled is selected. | Set of commands $C$ |
| 2 | The desired final sampling frequency $f$ and a temporal sampling duration $\tau$ are chosen | Sampling frequency $f$ and a temporal sampling duration $\tau$ |
| 3 | The exposure delay is set to $t=0$ | $t=0$ |
| 4 | The differential sampling mechanism in Table 11 (steps 2 to 12) are executed. | Set of images collected at temporal delay=0 |
| 5 | The exposure delay is incremented by $1/f$. | $t=t+1/f$ |
| 6 | Steps 4 and 5 are repeated $f/\tau$ times. | $f/\tau$ group of images |

Table 32

The goal of the test is to measure the optical amplitude of each command versus time, the command overshoot and the static error. For each of the command, a reference image is requested to compute the actual command amplitude. Such template may be given by the nominal command image (this stands for instance for the case of the mirror modes images); alternatively, all the realizations collected in the steady state may be averaged together, thus also reducing the effects of the steady state error.

**Parameter list: optical step responses**

| Parameter | Value | Description |
| --- | --- | --- |
| $f$ | 10kHz | final sampling frequency |
| $t$ | 0 to 3ms, 0.1ms step | exposure delays |
| $\tau$ | 0 to 3 ms | Temporal sampling duration |
| $C$ | 1;5;10;20;40;80;100;150; 300;500;700;800;1000 | Mirror modes sampled |

Table 33

### 4.10.2 Result

In Table 34 we present the procedure for the processing of the images datasets in order to build up the mirror temporal response to the given commands and to derive the requested parameters. The result is shown in Figure 22.

**Automatic analysis procedure: optical step responses**

| Step | Action | Result |
| --- | --- | --- |
| 1 | The $n$ template images are collected together in a matrix R | R: $n \times p$ matrix, where $p$ is the number of valid pixel within the mask. |
| 2 | R is pseudo inverted to obtain a projection matrix W | W= $R^{-1}$ |
| 3 | The dataset corresponding to the time delay $t$ is loaded. | |
| 4 | Each image $s_i$, corresponding to the mode $i$ within the dataset is projected with W. | Optical command amplitude $a_{it}$, associated to mode $i$ and delay $t$. |
| 5 | Steps 3 and 4 are repeated for all the datasets. | |
| 6 | The plot $a_{it}$ vs $t$ is drawn. | See figure |
| 7 | The settling time is measured as the delay when the measured amplitude is 90% the command | Settling time |
| 8 | The overshoot is measured on the plot | Overshoot |
| 10 | The steady state error is computed as the standard deviation of the amplitudes $a_{it}$ for delays larger than the settling time. | Steady state error |

Table 34

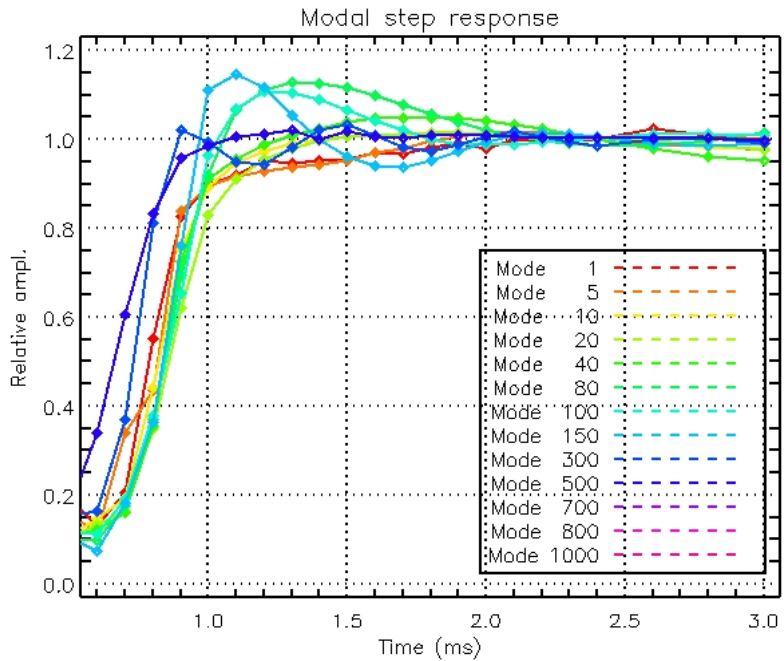

Figure 22 Optical step response, measured for a set of modes.

| Item | Value | Mode |
|---|---|---|
| Fastest settling time | 0.8 ms | 500 |
| Longest settling time | 1.0 ms | 20 |
| Largest overshoot | 15% | 150 |
| Largest steady state error | 6% | 1000 |
| Lowest steady state error | 0.5% | 500 |

Table 35 Optical step response results.

## 4.11 Automatic overnight measurements

The functionalities of the DSM may be demonstrated with overnight, unattended measurements, simulating a *seeing limited* observing run (i.e. no adaptive correction). In such scenario, the mirror is set to the default flat position which must be preserved throughout the observations. Although the largest part of the external working conditions cannot be modified (e.g. zenith angle, wind speed, telescope tracking), the test is useful to investigate the system reliability during a long lasting work load. On purpose, the cooling temperature may be modified to investigate the system response to temperature variations.

We considered two main outputs of the test: the number of recoverable and unrecoverable faults during the runs and the stability of the system parameters versus time and temperature variations. The test was performed approximately 60 times, 10 of which being full overnight runs.

### 4.11.1  Data sampling

In order to start the automatic procedure, the system must be fully thermalized at the working conditions (i.e. shell set and cooling temperature set to the desired value). In the case of a non constant temperature test, the cooling set point will be modified after measuring the initial conditions, for instance 1 or 2 hours after the start of the sampling procedure. the temperature set point may be modified according two different strategies: in a single step, so that from the step response one can derive the time behavior of the system, or following the maximum temperature rate specified for the observing conditions.

According to the definition of the test outputs defined above, the data to be collected are: mirror WFE, cold plate temperature, reference body temperature, electronic boards temperature, actuator position and forces, coolant temperature and flow, mechanical structure temperatures and HP position values.

The system will be maintained in optical alignment with the HP, with a low correction rate of (e.g.) one in 15 minutes. The data sampling, on the opposite, must be quite faster to allow an efficient averaging during the post-processing: this is for instance to reduce the convection noise, summing up images collected over a longer time span. The maximum frame rate must be however non faster than the convection coherence time.

Data will be stored in a dedicated folder according to the general structure defined in 2.5: within that folder, each individual file will be saved in file whose names are assembled with a prefix given by the tracking number of the current sampling and a suffix given by the data collected. For instance, the folder (see 2.5.2) /TimeSeries/20130101_100000/ will contain the following files:

/20130101_100100_opd.fits;

/20130101_100100_position.fits

/20130101_100100_temperatures.fits

/20130101_100200_opd.fits ….

Such sub-tracking number organization is intended to preserve time series structure of the sampling, allowing also to group files according to acquisition time.

**Automatic sampling procedure: time series measurements**

| Step | Action | Result |
|---|---|---|
| 1 | The system is powered on and the shell is set. | system ready |
| 2 | The system temperatures are monitored to check if thermalization has occurred. | |
| 3 | When the system is fully thermalized the sampling procedure is started. | |
| 4 | The sampling rate $s$ and alignment rate $a$ are selected. | sampling rate $s$ and alignment rate $a$ |
| 5 | An interferogram image is collected, averaging together n frames. The image is saved | image |
| 6 | The desired system parameters are measured and saved | system parameters measured. datafile |
| 7 | Steps 5 and 6 are repeated $s/a$ times | |
| 8 | The HP is activated to align the system. | HP aligned |
| 9 | The HP position values are read and saved | HP position saved |
| 10 | Steps 5 to 9 are repeated until the end of the test | |

Table 36

### 4.11.2 Result

The first test result is the full reliability of the DSM during the test: no failures were detected and the mirror was kept in seeing limited mode for the entire duration of the tests.

As a second point, we analyzed the system parameters and the mirror shape during the test, according to the procedure given in Table 37. The goal of the analysis is to monitor the DSM figuring versus time and to check the low order deformations coming from the cooling temperature variation.

**Automatic analysis procedure: time series measurements**

| Step | Action | Result |
|---|---|---|
| 1 | The dataset is selected | |
| 2 | The image files are listed (e.g. xxx_opd.fits) | Image files list. |
| 3 | A suitable averaging time $t$ is selected (e.g. 5 minutes) | |
| 4 | The images collected during the first $t$ are loaded and averaged together. | Average image over time $t$. |
| 5 | Alignment aberrations are computed and subtracted. Their amplitude is stored. | Alignment modes at time $t$ |
| 6 | The WFE of the average image is computed. | Aligned image WFE: $WFE_{aZ}$ |
| 7 | Mechanical aberrations (astigmatism and trefoil) are computed and subtracted. Their amplitude is stored. | Mechanical modes at time $t$ |
| 8 | The WFE of the average image is computed. | WFE after removing mechanical aberrations: $WFE_{mZ}$ |
| 9 | Steps 4 to 8 are repeated for all the images. | |
| 10 | The time series of $WFE_{aZ}$ and $WFE_{mZ}$ is plotted | WFE vs time |
| 11 | The time series of the fitted aberrations are plotted | Zernike vs time |

Table 37

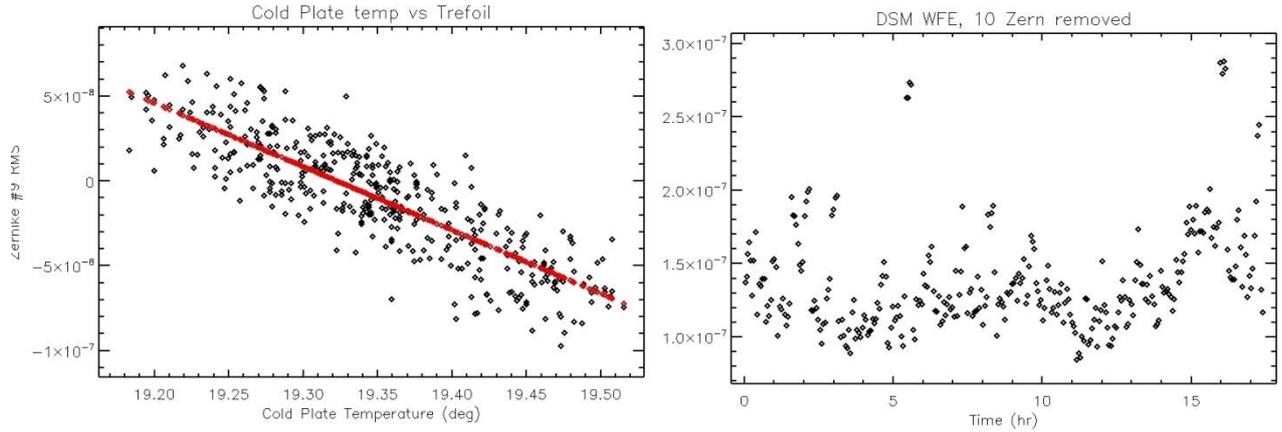

Figure 23 Left panel: Zernike Z6 trefoil versus cold plate temperature; right panel: WFE (rms) during the test, after the correction of 10 Zernike modes to compensate for thermo-mechanical drifts.

## 4.12 Noise evaluation

Within this scope we are not interested in a general and complete characterization of the noise in the calibration process, rather in the evaluation of its effect in the specific measurement cases. In Sec. 0 we identified two different types of measurements, involving differential or absolute sampling: each type has its own intrinsic noise, whose amplitude depends on the measurement parameters.

### 4.12.1 Differential sampling noise

The case of differential sampling is applicable in the following measurements:
- mirror IF, when each modal shape is sampled with opposite amplitude on n frames, which are later composed together to generate the IF image;
- capacitive sensor calibration, when at each step of the procedure an opposite trefoil command is applied and sampled;
- field stabilization verification, when two opposite tilt command are applied and sampled sequentially to measure wavefront departure from a pure tilt of the command.

The three procedures operate with the same sampling mechanism: a sequence of mirror commands of opposite amplitude is applied and sampled with the interferometer; the frame rate may be different depending on the case, it is however fast (15-25 Hz, e.g.). The images sequence (composed by $n$ frames) is then combined together using the equation given in 0, consisting in averaging all the differences between consecutive images. The residual noise in the sampling procedure is dependent on the frame rate value and $n$.

#### 4.12.1.1 Data sampling

For the measurement of the differential sampling noise, we run the procedure depicted above applying at each step a null command to the mirror. As a result, the analysis procedure yielded the residuals to be expected while collecting $n$ differential frames at the specified frame rate.

Such procedure may be used both for evaluating tip/tilt noise (for instance, to measure the level of vibration) or higher order noise (most likely corresponding to convection noise, tip/tilt corrected). We also measured the high order WFE, after additionally removing astigmatism and trefoil, which may be drift contributor at the time scale of the sampling. The detailed procedure is described in Table 38 and the result is shown in Figure 24.

#### 4.12.1.2 Result

The dataset collected has been analyzed according to the procedure in Table 38; the results are shown in Figure 24.

**Automatic analysis procedure: noise propagation through differential sampling**

| Step | Action | Result |
|---|---|---|
| 1 | The mirror is commanded to "flat" state | |
| 2 | A large sequence of $n$ images ($n$=1000, e.g.) is sampled with the interferometer at the desired frame rate. | Dataset composed of $n$ images. |

| 3 | The dataset (sequence of frame) is stored with the proper tracking number. | Folder containing the dataset. |
|---|---|---|
| 4 | A set $p$ of images sequence length is selected, corresponding to a vector of number-of-frames to be combined together (e.g.: 3,5,7,11,15,21,35,51). | |
| 5 | The images sequence is analyzed according to eq XXX: every $p_i$ frames a combined image is produced. | Sample of $n/p_i$ resulting images. |
| 6 | The tip/tilt RMS amplitude of each combined image is computed. | Sample of $n/p_i$ tip/tilt values. |
| 7 | The mean value and standard deviation of the tip/tilt sample is computed. | |
| 8 | The astigmatism RMS amplitude of each combined image is computed. | Sample of $n/p_i$ astigmatism values. |
| 9 | The mean value and standard deviation of the astigmatism sample is computed. | |
| 10 | Each of the combined image is tip/tilt and focus corrected. | $n/p_i$ resulting images, alignment corrected. |
| 11 | The WFE RMS of each combined image is computed | Sample of $n/p_i$ WFE RMS values. |
| 12 | Each of the combined image is astigmatism and trefoil corrected. | $n/p_i$ resulting images, low order modes corrected. |
| 13 | The WFE RMS of each combined image is computed. | Sample of $n/p_i$ WFE RMS values. |
| 14 | The mean value and standard deviation of the RMS sample is computed. | |
| 15 | Steps 5 to 14 are repeated changing the $p_i$ parameter. | |
| 14 | In the end, the plots $p$ vs tip/tilt and $p$ vs WFE are obtained. | |

Table 38

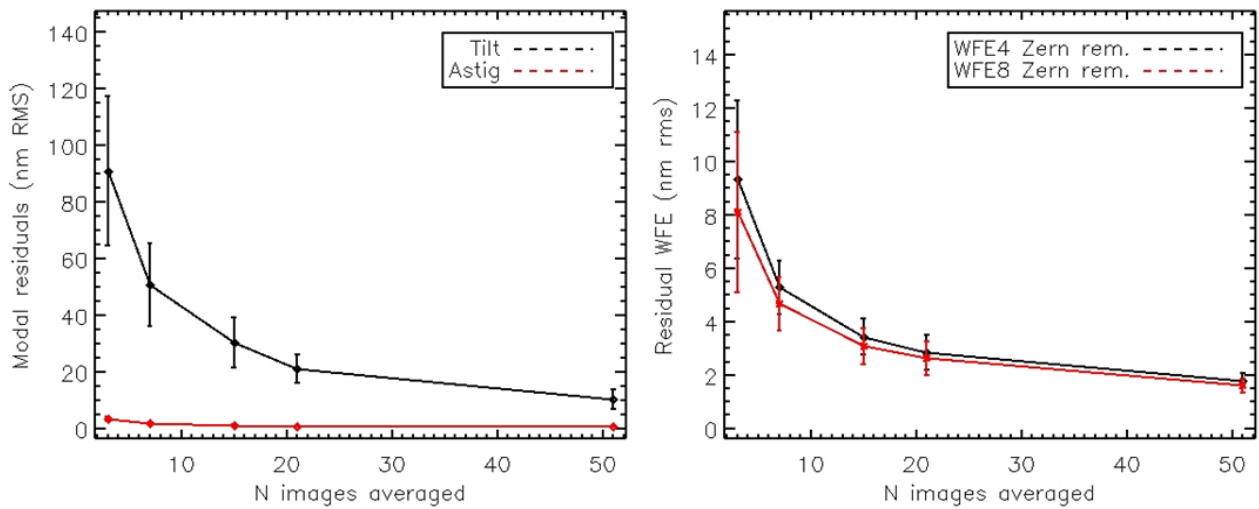

Figure 24 Left panel: residual tip/tilt ans astigmatsim, after averaging together $n$ differential images; right panel, residual WFE (RMS), after averaging together $n$ differential images.

### 4.12.2 Absolute sampling

The case of the absolute sampling is applicable in the following measurements:
- Mirror flattening and result verification;
- Last stage of the capacitive sensor calibration;
- Field stabilization verification (large stroke).

Absolute measurements are done by averaging together $n$ frames collected over a time span $t$. We invoke here the assumptions that the $n$ samples are statistically uncorrelated (i.e. the sampling period $t/n$ must be larger than the noise coherence time $\tau$) and that the noise is a stationary process, at least for a time $t_0$. Under these assumptions, the measurement noise is $\sigma_\infty/\sqrt{n}$, where $\sigma_\infty$ represents the noise expected for a single shot measurement.

### 4.12.2.1 Data sampling

$\tau$ and $\sigma_\infty$ are measured by computing the noise structure function. The starting dataset is the same for the IF measurements (with no command applied) as described in Table 38 and will sampled following the same procedure. By changing the acquisition frame rate, it is possible to compute the structure function at different time scales and eventually verify the noise stationarity. The coherence time is defined as the "knee" of the structure function; $\sigma_\infty$ is $\sqrt{2}$ times its asymptotic value. The detailed procedure is given in Table 39.

Non stationarity (at least at the typical time scale of the calibration process, 8hr e.g.) may be evident by observing quasi-periodic or slowly varying phenomena and long term drifts. Such unfavorable conditions are very habitual in the testing environment, and are given for instance by cooling temperature oscillations, thermo-mechanical drifts, day-night and human activity cycles. We estimated a typical $t_0$ of 15 minutes.

Once we estimated $\tau$, $t_0$, we performed absolute samplings by collecting $t_0/\tau$ frames at $1/\tau$ frame rate. A direct estimation of the associated noise is obtained by computing the WFE RMS of the difference between two absolute images (sampled as explained) separated by a time interval $t_0$, as will be described in Table 41. Such result may be then compared with the expected $\sigma_\infty$.

### 4.12.2.2 Result

The noise structure function is obtained by running the procedure depicted in Table 39. A typical realization of the noise structure function, in the time interval 0s to 18s, is given in Figure 25.

**Automatic analysis procedure: noise structure function**

| Step | Action | Result |
|---|---|---|
| 1 | A dataset of *n* images sampled at frame rate *f* is collected. | Images dataset |
| 2 | The dataset is saved with the proper tracking number. | Dataset file |
| 3 | A set *v* of time delay values is chosen. The number *b* of points to average together is selected (e.g. 20). | Time delay vector |
| 4 | The first value of time delays *v* is considered, corresponding to a frame index offset of *vf* | |
| 5 | The images *i+vf* and *i+2vf* within the sequence are loaded. | |
| 6 | The difference between them is computed and tip/tilt corrected. | Differential images, at *v* time offset. |
| 7 | The WFE RMS is computed and stored. | WFE of the differential images. |
| 8 | Steps 5 to 7 are repeated *b* times. | |
| 9 | The WFE values are averaged together | Mean WFE of the differential images. |
| 10 | Steps 5 to 8 are repeated, for each of the *v* values. | |
| 11 | The curve WFE vs *v* is plotted, representing the square root of the noise structure function. | Plot: WFE vs *v* |
| 12 | The coherence time and asymptotic value are measured on the curve. | $\sigma_\infty$ and $\tau$ |

Table 39

**Parameter list: noise structure function**

| Parameter | Value | Description |
|---|---|---|
| *v* | 0.05s, 0.1s, 0.3, 0.5s, 1s, 3s, 5s, 10s, 15s | time delays |
| *b* | 20 | number of differential frames to be averaged together |
| *f* | 25 | frame rate |

Table 40

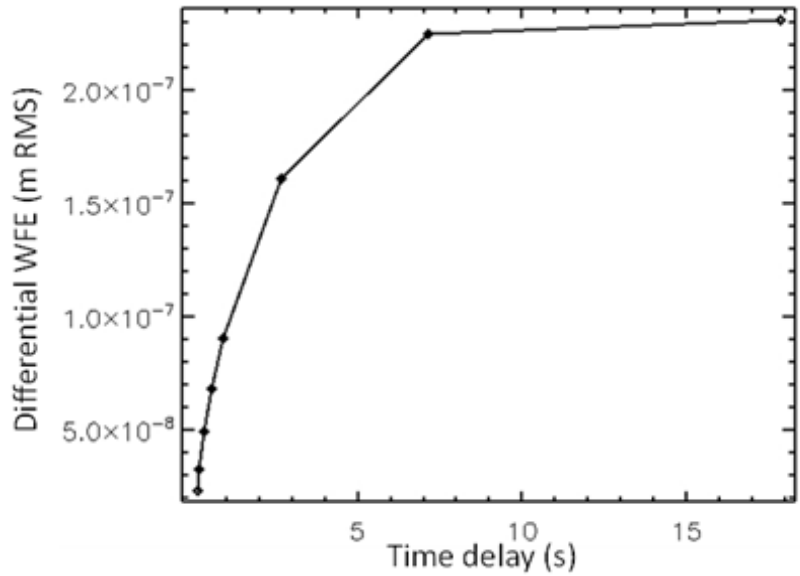

Figure 25 Square root of the noise structure function, computed according the procedure described above. The coherence time is estimated to be 5s and the asymptotic value 250 nm RMS.

**Automatic analysis procedure: absolute sampling noise**

| Step | Action | Result |
|---|---|---|
| 1 | $\tau$, $t_0$ have been measured or estimated. | Noise coherence and stationarity time |
| 2 | A first image is sampled by averaging $t_0/\tau$ frames at $1/\tau$ frame rate. | first averaged image |
| 3 | A second image is sampled with no time delay after the first; it is sampled by averaging $t_0/\tau$ frames at $1/\tau$ frame rate. | second averaged image |
| 4 | Both images are corrected for tip/tilt and focus. | images, alignment corrected |
| 5 | The intersection mask is computed. | |
| 6 | The difference between them is computed. | differential image |
| 7 | The RMS within the intersection mask is computed. | WFE RMS |

Table 41

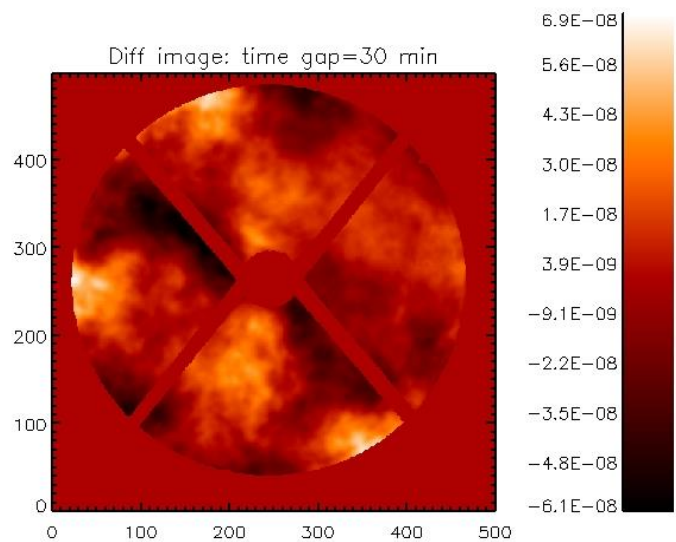

Figure 26 The differential image computed according the procedure. The WFE is 44 nm RMS.

# 5   Last remark and conclusions

Measuring and calibrating a large format deformable mirror requires a dedicated test setup, including an interferometric bench, alignment stages and a control/interfacing software. The complexity and cost of the setup scale with the mirror dimensions, on the one hand, and with the performance or accuracy requirements, on the other. All these elements shall be put into a trade-off matrix, in order to keep the measurement system within given complexity and cost boundaries.

The interferometer is the core of the measurement process. In Sec. 3.2 we summarized the functional and operational needs for the interferometer but we did not discuss specifically its set-up or the selection of the working parameters. Since they are dependent from the interferometer model/type, the reader may refer to the manufacturer handbook. Some recent papers are listed in the references.

The environment is another key-actor and the measurement accuracy and precision depends largely on the sampling conditions: for instance, the presence of external walls close to the optical path, or windows/roofs may have strong impact on convection noise; while vibration noise may be injected by human activities nearby or even traffic. In Sec. 4.12 we described how to assess the measurement noise and to select properly the sampling parameters. In general, a valuable strategy is to turn the environment noise into a known disturb, in order to set optimal sampling parameters: for instance, the convection noise may be "standardized" by means of fans.

As a last remark, we point out that the field of large format adaptive mirrors is in its golden age: after the successful deployment of the LBT, Magellan and VLT adaptive secondaries, now it's the turn of the wavefront correctors for the extremely large telescopes, such as M4 for the ELT and the seven adaptive M2s for the GMT. What's next? Large format secondaries have been proposed as deformable primaries (or segments of a primary) for the next generation space telescopes. In these areas, new challenges are showing up in the context of optical metrology: the co-alignment and co-phasing of the segments, the replacement of defective actuators, the fully-automated calibration and check on a compressed schedule. New strategies and procedures shall be developed to face these challenges!

# Acknowledgments

This work has been carried out under an INAF (Italian Institute or Astrophysics) grant, TECNO-INAF 2010.
The authors acknowledge the support of their colleagues in the Adaptive Optics group at the Arcetri Observatory and of the ADONI community (Adaptive Optics National laboratory Italy).

The authors wish to thanks their colleagues and collaborators at ESO and AdOptica (ADS-International and Microgate) for the stimulating, challenging and successful optical calibration activities on the DSM at ESO HQ in Garching.